\documentclass[aps,pre,twocolumn,eqsecnum,amsmath,floatfix,showpacs]{revtex4-1}
\usepackage{epsfig}
\usepackage{times}
\usepackage[sf]{subfigure}
\usepackage{bm}
\usepackage{longtable}

\newcommand{\diag}[1]{\scalebox{0.13}{\mbox{\epsfig{file=graph#1.eps}}}}
\renewcommand{\atop}[2]{\genfrac{}{}{0pt}{}{#1}{#2}}
\begin{document}

\title{Generalized two-body self-consistent theory of random linear dielectric composites:\\
an effective-medium approach to clustering in highly-disordered media}
\author{Y.-P.~Pellegrini}
\affiliation{ CEA, DAM, DIF, F-91297 Arpajon, France}
\author{F.~Willot}
\affiliation{MINES Paritech, Centre de Morphologie Math\'ematique, Math\'ematiques et Syst\`emes, 35, rue Saint-Honor\'e, F-77305 Fontainebleau Cedex, France}
\date{\today}
\begin{abstract}
Effects of two-body dipolar interactions on the effective permittivity/conductivity of a binary, symmetric, random dielectric composite are investigated in a self-consistent framework. By arbitrarily splitting the singularity of the Green tensor of the electric field, we introduce an additional degree of freedom into the problem, in the form of an unknown ``inner'' depolarization constant. Two coupled self-consistent equations determine the latter and the permittivity in terms of the dielectric contrast and the volume fractions. One of them generalizes the usual Coherent Potential condition to many-body interactions between single-phase clusters of polarizable matter elements, while the other one determines the effective medium in which clusters are embedded. The latter is in general different from the overall permittivity. The proposed approach allows for many-body corrections to the Bruggeman-Landauer (BL) scheme to be handled in a multiple-scattering framework. Four parameters are used to adjust the degree of self-consistency and to characterize clusters in a schematic geometrical way. Given these parameters, the resulting theory is ``exact'' to second order in the volume fractions. For suitable parameter values, reasonable to excellent agreement is found between theory and simulations of random-resistor networks and pixelwise-disordered arrays in two and tree dimensions, over the whole range of volume fractions. Comparisons with simulation data are made using an ``effective'' scalar depolarization constant that constitutes very sensitive indicator of deviations from the BL theory.
\end{abstract}
\pacs{PACS numbers: 05.60.Cd, 72.80.Tm, 78.20.Bh}
\maketitle

\section{Introduction}
The problem of the effective transport properties of disordered media has a long history \cite{LAND78,BROS06}, and still continues to attract wide attention owing to its intrinsic theoretical interest \cite{TORQ01,MILT02,SAHI03}, and to its importance to various domains of engineering sciences \cite{RENA97,*MIKR99,*AMBR07}. Our focus here is on the effective permittivity of binary symmetric random media \cite{BROW55} with quenched disorder, that undergo percolative behavior \cite{LAND78,STAU94,*SAHI94}. Investigations have mainly been carried out on prototypical models such as Random Resistor Networks (RRN) \cite{STRA77,CLER90,CLER96}, random arrays of polarizable point elements \cite{CHEN91,*BART97,*SONG00} and checkerboards \cite{SODE83}. The advent of full-field numerical methods of computation, such as Fast Fourier Transform (FFT) calculations on pixel arrays \cite{MICH00,WILL08a} (see \cite{EYRE99,*MAZH00} for an alternative use of FFTs in this context), or finite-elements methods, make it now possible to address in detail random checkerboards and various other types of heterostructures \cite{MEJO06}.

From a theoretical standpoint, accounting for presence of a percolation threshold in a systematic theory of interactions between heterogeneities is extremely difficult, since the highly-disordered character of the percolation regime involves many-body positional correlation functions to all orders \cite{TORQ01,SAHI03}. There, perturbative methods are inapplicable unless some amount of self-consistency is injected in suitable approximations. The simplest successful \cite{BERG07} self-consistent (s.c.) approach to the effective permittivity problem in percolating media is the well-known Bruggeman-Landauer (BL) theory \cite{BRUG35,LAND78}, whose theoretical status is well-established with regard to perturbative approaches \cite{HORI75a,BERG81b,*LUCK91}. It can be applied to RRNs and to continuum systems. In the BL theory, as well as in more elaborate s.c.\ treatments \cite{HORI75a,SAHI83a,*PELL00}, the threshold is fixed, which is a nuisance since it varies in practice with microstructure \cite{SAHI03}.

The present work proposes a parametric theory to account for corrections to the BL theory in the high-contrast, highly-disordered regime. Simple ways of doing so mostly rely on varying the shape of the inclusions by modifying their depolarization coefficients \cite{GRAN78,*GONC03}. The related formulations of McLachlan and co-workers \cite{MCLA86,*WUMC97} empirically introduce tunable critical exponents and threshold in the BL effective-medium formula. In contrast, our purpose is to incorporate corrections to the BL theory by considering exact pairwise interaction terms between polarizable pointlike matter elements, which will be done in the generic continuum framework of Miller's cell-material model \cite{MILL69a,HORI75c}. In this connection, it should be mentioned that Shen and Sheng previously accounted for nearest-neighbor and next-nearest-neighbor interactions -- a special case of pairwise terms -- in a BL-like framework, by using two-dimensional split inclusions \cite{SHEN80b}, which resulted in a theory with a double percolation threshold \cite{SHEN82}. This double-threshold effect is nowadays actively discussed \cite{NETT03,*SNAR08}, notably in the context of random checkerboards \cite{CHEN09,*HELS11b,*HELS11a}.

Our starting point is the modified BL equation for the effective permittivity $\varepsilon_e$ \begin{equation}
\label{eq:brugellmod}
\left\langle\frac{\varepsilon-\varepsilon_e}
{\ell_e\varepsilon+(1-\ell_e)\varepsilon_e} \right\rangle=0.
\end{equation}
The brackets denote an average over material cells of variable permittivity $\varepsilon$, and $\ell_e$ is an unknown effective depolarization coefficient to be determined. In the original BL theory, $\ell_e=1/d$ where $d$ is the space dimension, which is equal to the percolation threshold. Eq.\ (\ref{eq:brugellmod}) arises in a variety of contexts. In particular, the Wu--McLachlan formula reduces to the above one when its exponents are set to one \cite{WUMC97}. Rather, our line of thought will be that $\ell_e$ is a dimensionless \emph{function} of permittivity ratios and of the volume fraction of phases \cite{PELL93,BALI95}.

The phenomenology of $\ell_e$ is established in Sec.\ \ref{sec:rsrg} by identifying the solution of Eq.\ (\ref{eq:brugellmod}) with results of simulations of binary RRN and pixelwise-disordered systems in two and three dimensions. In Sec.\ \ref{sec:mulscatfor}, we develop a formalism rooted in multiple-scattering theory, in which a free, inner, depolarization constant $\ell$, analogous to the above effective one, can be introduced without making approximations. Independently, microstructure-related features are approximately accounted for by means of two parameters, which provides a rough characterization of single-phase clusters. With this simple device, we make an bypass the need to consider explicitly correlation functions. Section \ref{sec:emconds} formulates the \emph{two} s.c.\ conditions required to determine $\ell$ and the effective permittivity. One of these modifies the usual \emph{coherent-potential condition} of vanishing self-energy, into one that distinguishes between the local and non-local parts of the two-body term in the self-energy. It serves to adjust, by means of two supplementary parameters, the ``amount of locality and non-locality'' that enter the s.c. condition. This peculiar treatment is justified by comparing the resulting four-parameter theory with the simulations of Sec.\ \ref{sec:rsrg}. We conclude in Sec.\ \ref{sec:concl}.

\section{Phenomenological behavior of $\ell_e$}
\label{sec:rsrg}
\subsection{Preliminary remarks}
\label{sec:prelrem}
We consider a $d$-dimensional binary medium, with phases of  permittivities $\varepsilon_1$ and $\varepsilon_2$ in respective proportions $1-f$ and $f$. For definiteness, we choose phase 2 as that of high permittivity. Eq.\ (\ref{eq:brugellmod}) then reduces to the second-order polynomial equation for  the effective permittivity $\varepsilon_e$
\begin{equation}
\label{embin} (1-f)\frac{\varepsilon_1-\varepsilon_e}
{\ell\varepsilon_1+(1-\ell_e)\varepsilon_e}+f\frac{\varepsilon_2-\varepsilon_e}
{\ell\varepsilon_2+(1-\ell_e)\varepsilon_e}=0,
\end{equation}
In the dilute limit, an exact well-known \cite{SAHI03,SNAR07} result (due to Maxwell in three dimensions) is that:
\begin{equation}
\label{exactdof}
\varepsilon_e/\varepsilon_1=1+f\frac{d(\varepsilon_2-\varepsilon_1)}
{\varepsilon_2+(d-1)\varepsilon_1}+O(f^2).
\end{equation}
On the other hand, expanding (\ref{embin}) provides:
\begin{equation}
\label{ellof}
\varepsilon_e/\varepsilon_1=1+f\frac{\varepsilon_2-\varepsilon_1}
{\ell\varepsilon_2+(1-\ell_e)\varepsilon_1}+O(f^2),
\end{equation}
where $\ell_e$ stands for $\ell_e(f=0)$. The $O(f)$ term arises from the exact polarizability factor that individually characterizes the impurities (one-body term). Thus, $\ell_e(f=0)=1/d$ and the $O(f)$ correction to this value is due to the $O(f^2)$ term in $\varepsilon_e$, which accounts for pairwise (two-body) interactions. Similar considerations hold near $f=1$ where $\ell_e(f=1)=1/d$.

Next, percolative behavior takes place when $\varepsilon_1\ll\varepsilon_e\ll\varepsilon_2$. Letting $\varepsilon_e/\varepsilon_1\to\infty$ and $\varepsilon_2/\varepsilon_1\to\infty$ at the percolation threshold $f=f_c$ in (\ref{embin}) provides the following equation for $f_c$ \cite{PELL93}:
\begin{equation}
\label{percothr}
\ell_e(f_c)=f_c.
\end{equation}
In the critical region, differences are expected between two- and three-dimensional cases. For infinite contrast, the modified BL model (\ref{embin}) reduces to $\varepsilon_e=\varepsilon_1(1-f/\ell_e)^{-1}$ for $f<f_c$, and $\varepsilon_e=\varepsilon_2(f-\ell_e)/(1-\ell_e)$ for $f>f_c$. On the
other hand, the well-known critical behavior of binary media in the
infinite-contrast limit is \cite{SAHI03}
$\varepsilon_e\propto\varepsilon_1(1-f/f_c)^{-s}$ for $f<f_c$ and
$\varepsilon_e\propto\varepsilon_2(f/f_c-1)^t$ for $f>f_c$, with
$d$-dependent critical exponents $s$ and $t$. Comparing these
equations provides
\begin{subequations}
\begin{eqnarray}
\ell_e(f)&\simeq&f+a^-f_c(1-f/f_c)^s\quad\qquad(f\lesssim f_c),\\
       &\simeq&f-a^+(1-f_c)(f/f_c-1)^t\quad (f\gtrsim f_c).
\end{eqnarray}
\end{subequations}
where $a^{\pm}>0$ are coefficients of order one. For $d=2$, exponents are such $s=t\simeq 1.3>1$, so that derivatives at $f_c$ are
\begin{equation}
\label{eq:derivs2}
\ell_e'(f_c^\pm)=1\qquad (d=2),
\end{equation}
and $\ell_e(f)$ is smooth. For $d=3$ instead, exponents are now $s<1$ and $t>1$ and
\begin{equation}
\label{eq:derivs3}
\ell_e'(f_c^-)=-\infty,\qquad \ell_e'(f_c^+)=1\qquad (d=3),
\end{equation}
so that $\ell_e(f)$ must display a cusp at $f_c$.

These considerations must be modified when applied to a mean-field theory such as the one developed hereafter, in which the critical exponents would have typical effective-medium values $s=t=1$ irrespective of $d$. Then,
\begin{subequations}
\begin{eqnarray}
\label{eq:derivemm}
\ell_e'(f_c^{-})&=&1-a^{-},\\
\label{eq:derivemp}
\ell_e'(f_c^{+})&=&1+a^+(1-f_c^{-1}).
\end{eqnarray}
\end{subequations}
Depending on the values of $a^\pm$ and $f_c$, $\ell_e'(f_c^\pm)$ can then (a priori) be of either sign.

To close, we point out that the double-threshold effect for two-dimensional checkerboards (see Introduction), translates in terms of $\ell_e$ into the property $\ell_e(f)\simeq f$ for
$f\in(p_c,1-p_c)$, where $p_c$ is the two-dimensional site-percolation threshold \cite{SHEN82}.

\subsection{Simulation data}
\label{ref:refdat}
\begin{figure*}[!ht]
\centering
\includegraphics[width=4.3cm]{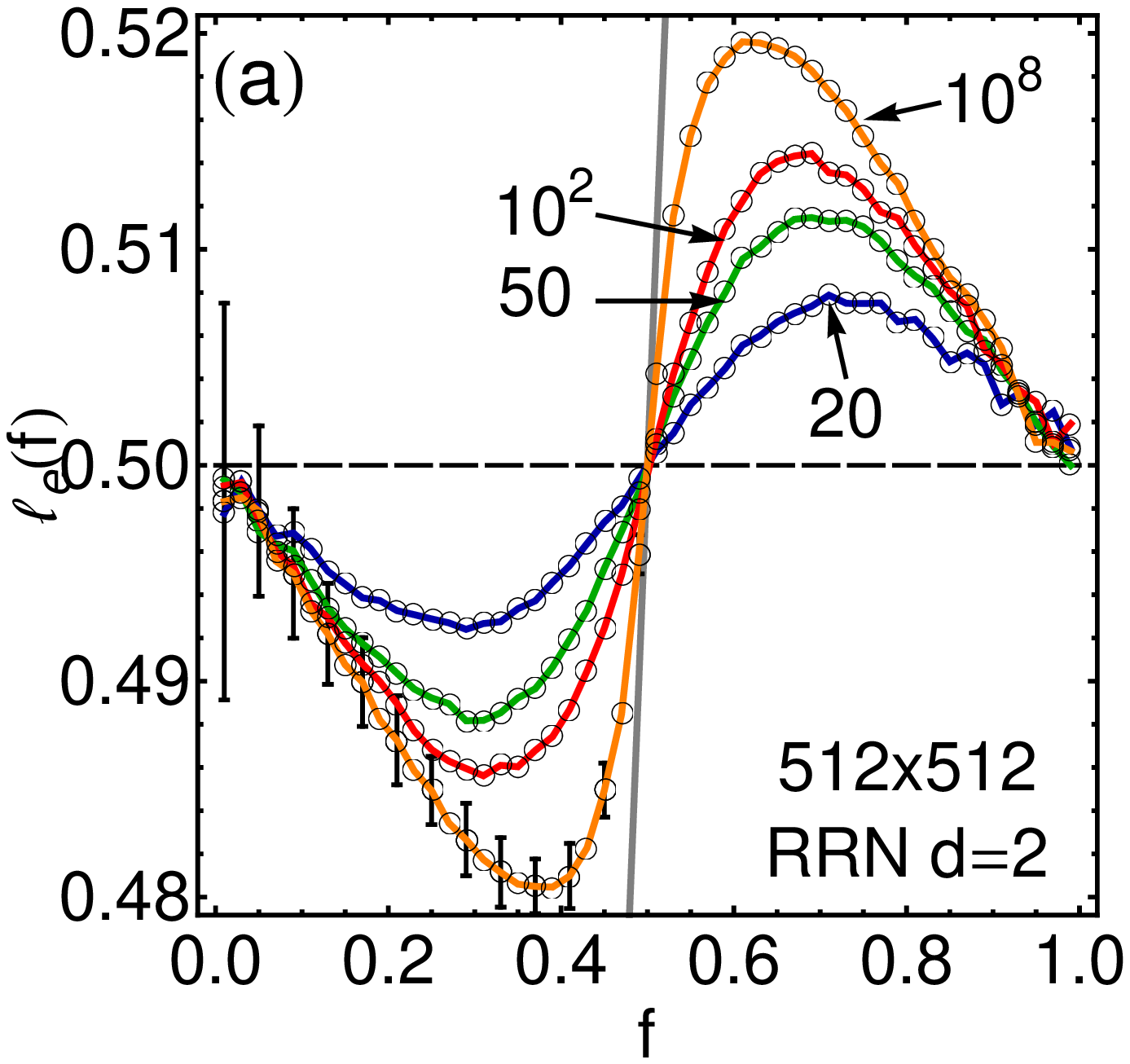}
\includegraphics[width=4.3cm]{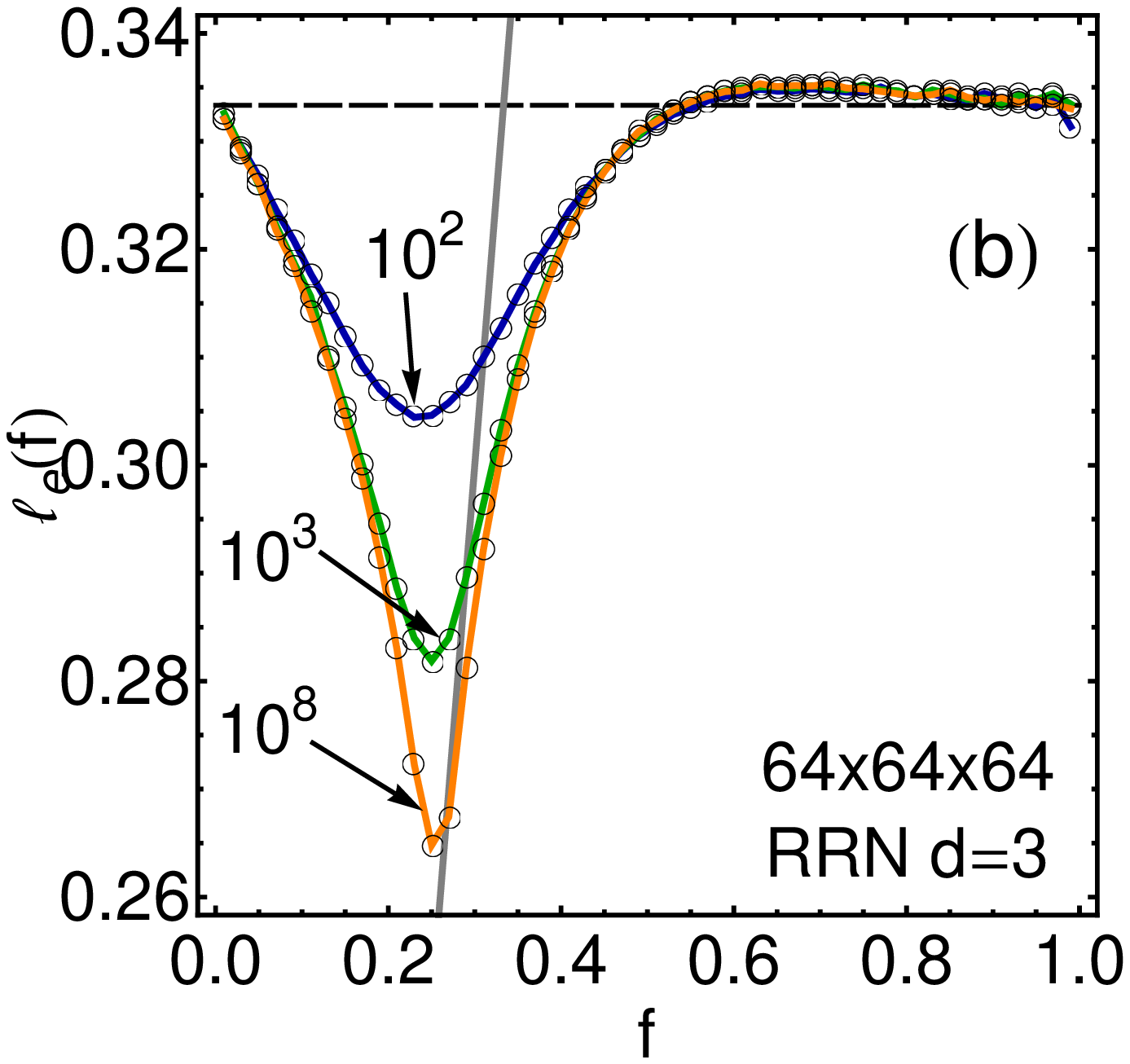}
\includegraphics[width=4.3cm]{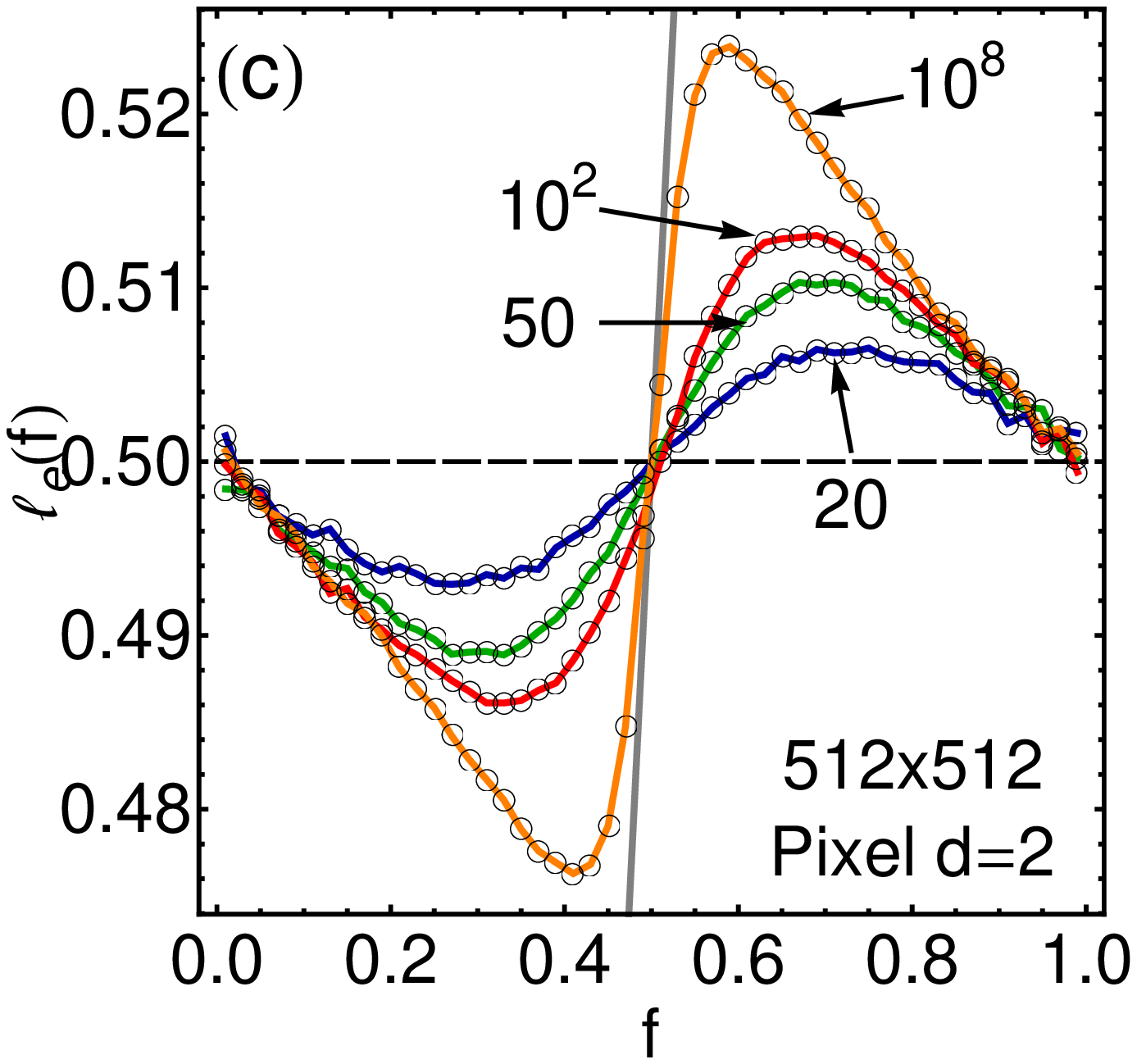}
\includegraphics[width=4.3cm]{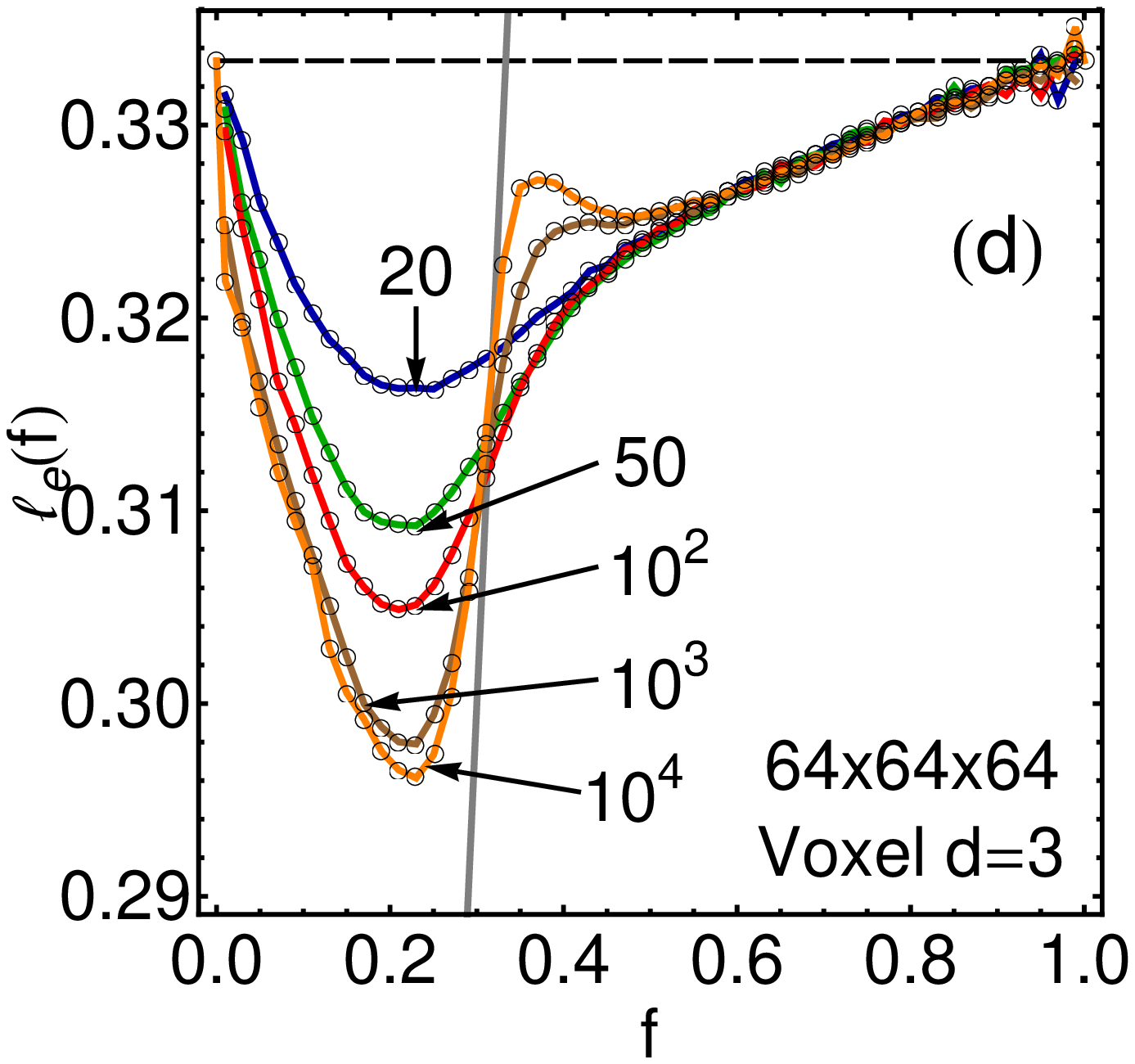}
\caption{\label{fig:fig1} (Color online) Permittivity (conductivity) data analyzed in terms of $\ell_e(f)$ for two- and three-dimensional models, and contrasts $\varepsilon_2/\varepsilon_1$ as indicated in the plots: (a) and (b), random-resistor networks of size $L=512$ (resp., $L=64$) for $d=2$ (resp., $d=3$); (c) and (d), pixelwise-disordered media of same sizes (see text). Dashed: $1/d$. Quasi-vertical grey line: function $\ell_e=f$. Solid lines between data points are guides to the eyes.}
\end{figure*}
The BL theory compares well to binary RRNs, provided that an identification is made
between the phase volume fraction and the volume fraction of bonds,
because both are symmetric with respect to phase interchange
\cite{KIRK71,LUCK91}. In particular, for $d=2$ the percolation threshold
coincide, which leads to an overall reasonable match.
Moreover, the effective conductivity (permittivity) in the BL theory coincides
with that of RRNs up to the third order (included) of perturbations
in powers of the scaled contrast $\delta
\varepsilon/\langle\varepsilon\rangle$ \cite{LUCK91}, and it is
exact to one-body order (included) in the sense of Sec.\
\ref{sec:ensav}. However, strong discrepancies
with RRNs arise for $d\geq 3$, primarily because the
percolation threshold of the BL theory ($f_c=1/d$) overestimates
that of the lattice system.

Discrepancies can be analyzed by expressing $\ell_e(f)$ as a function of $\varepsilon_e$ by means of (\ref{embin}), as \footnote{This equation is closely related to Eq.\ (13) of Soven \cite{SOVE67}.}
\begin{equation}
\label{eq:elleps}
\ell_e=\frac{\varepsilon_e(\langle\varepsilon\rangle-\varepsilon_e)}{(\varepsilon_e-\varepsilon_1)(\varepsilon_2-\varepsilon_e)},
\end{equation}
and by using for $\varepsilon_e$ effective-permittivity data obtained from models. We consider three numerical models.

To begin with, Fig.\ \ref{fig:fig1}(a) and (b) show an analysis of the effective conductivity of simulations of bond-disordered RRNs, carried out for the present purpose. The function $\ell_e(f)$ is computed from (\ref{eq:elleps}) for various contrasts. Statistical averages of the overall conductivity of a sufficiently large number of samples have been carried out to reduce standard deviations to the typical error bar values represented in Fig.\ \ref{fig:fig1}(a) for $f<1/2$. Although not represented, they are of same order of magnitude in the domain $f>f_c$, and in Figs.\ \ref{fig:fig1}(b), (c) and (d). Under-sampling of the dilute configurations makes the statistical errors larger in the limits $f\to 0,1$.  System sizes are large enough to make finite-size effects negligible to our purpose. According to (\ref{percothr}), $f_c$ is the volume fraction $f$ at the crossing point between the infinite-contrast plots and the grey line that represents $\ell_e=f$. It is seen that $f_c(d=2)=1/2$, and $f_c(d=3)\simeq 0.26$, close to the expected theoretical bond-percolation value $\simeq 0.249$ . The dilute-limit value $\ell_e=1/d$ is represented by the dashed horizontal line, and is approached with \emph{negative} slopes in all cases. Note that in three dimensions $\ell_e$ slightly \emph{exceeds} $1/3$ for $f\gtrsim 0.55$. In this high-conductivity region, the closeness to Bruggeman's theory is remarkable.

We also carried out noiseless calculations using Bernasconi's Real-Space Renormalization-Group model for RRNs \cite{BERN78}. Up to some limitations of the approach, and to a lower percolation threshold in three dimensions, results are good qualitative agreement with Figs.\ \ref{fig:fig1}(a) and (b) \cite{*[{}] [{See supplementary material at \url{http://link.aps.org/supplemental/xx.xxxx/PhysRevE.XX.xxxxxx}}] EPAPS}.

Figs.\ \ref{fig:fig1}(c) and (d) display further results of simulations on pixelwise-disordered arrays (PDA) \cite{WILL08a}, solved using the FFT method developed for elastic composites by Michel et al.\ \cite{MICH00}. Its adaptation to linear dielectric media is straightforward. The formulation employed is the original one, that uses the continuum Green function (other types of implementation were considered in \cite{WILL08a}). Each sample is a regular square or cubic array of $L^d$ pixels (voxels, in three dimensions), of permittivity chosen at random according to the binary probability density. Conceptually closer to a random array with substitutional disorder, than to a bond network or a random checkedboard (the fields are not resolved within the pixels or voxels), this system nonetheless features in two dimensions the same percolation threshold $f_c=1/2$ as a bond-disordered RRN; for $d=3$, $f_c\simeq 0.315$, a value reminiscent of site percolation ($f_c\simeq 0.312$), close to the Bruggeman value. For $d=2$ its $\ell_e(f)$ function resembles that of RRNs, with a sharper variation at threshold. The situation changes in three dimensions, especially in the high-permittivity phase where deviations from the dilute limit markedly differ from that in RRNs: the approach of $f=1$ has a \emph{positive} slope; the concavity at small $f$ is opposite; moreover, in the infinite-contrast limit, $\ell_e(f)$ develops a valley just after the percolation threshold. The cusp at $f\simeq 0.36$ marks out a transition from the critical region where Eq.\ (\ref{eq:derivs3}) (right) applies, to behavior of the effective-medium  type (\ref{eq:derivemp}). The critical region $f\lesssim f_c$ where Eq.\ (\ref{eq:derivs3}) (left) would lead to a cusp  is not observed. This suggests that the critical behavior of this system, which has no contact interactions, might be different from the RRN one, with either $s\simeq 0$, or $s=1$ with $a^-\ll 1$. This point --not crucial to our purpose-- has not been investigated further, due to some difficulties in achieving high-quality numerical convergence for $f<f_c$ in the infinite-contrast limit (this is the reason why contrast is limited to $10^4$).

The rich typology of $\ell_e(f)$ behaviors, even in the dilute region, illustrates the dramatic influence of microstructural features on the overall response. This makes $\ell_e(f)$ an interesting means of analysis, since plots of $\ell_e$ provide nontrivial information over the whole range of concentrations (trying to use it at low contrast on noisy data may however lead to an inconclusive outcome \cite{PELL93}).

\section{Theory}
\label{sec:mulscatfor}
The theoretical framework we adopt to account for the above observations heavily relies on the well-known multiple-scattering formalism \cite{FOLD45,*LAX51,*LAX52,*WATE61,FRIS68}, widely used in solid-state physics \cite{EBER11}, and repeatedly employed to study dielectric media \cite{BARR89,GARC99,ZAKH12}. Because some additions to the classical setup are needed, the following sections review it briefly with emphasis on our modifications. A number of the equations also arise in the context of dilute alloys, a related problem, where Bruggeman's EMA is known as the \emph{Coherent-Potential Approximation} (CPA) \cite{SOVE67}; see \cite{ROWL09} for a recent review.
\subsection{Split-up of the Green function}
\label{sec:splitup}
We consider a $d$-dimensional dielectric medium with permittivity $\varepsilon(\mathbf{r})$ fluctuating from
cell to cell. The latter are spherical on average, of infinitesimal typical radius $a\to 0$ and
volume $v=a^d S_d/d$, where $S_d=2\pi^{d/2}/\Gamma(d/2)$ is the area of the $d$-dimensional sphere of unit radius. We use the notations $\mathsf{I}$ for the $d\times d$ identity matrix, and $\mathbf{\hat r}=\mathbf{r}/r$ for the unit position vector. Following the usual treatment, we introduce an arbitrary background permittivity $\varepsilon_b$, and re-write the equilibrium equation $\nabla.(\varepsilon\mathbf{E})=0$,
where $\mathbf{E}$ is the electric field, as $\varepsilon_b\nabla.\mathbf{E}=-\nabla.\delta\varepsilon\mathbf{E}$,
where $\delta\varepsilon=\varepsilon-\varepsilon_b$ is the dielectric contrast. The problem can then be cast in the form of an integral equation for $\mathbf{E}$ \cite{STRO75}:
\begin{equation}
\label{inte}
\mathbf{E}(\mathbf{r})=\mathbf{E}_0(\mathbf{r})+\int
\mathrm{d\!}^d\!r'\,\mathsf{G}_{1/d}(\mathbf{r}-\mathbf{r}')\frac{\delta\varepsilon(\mathbf{r}')}
{\varepsilon_b}\mathbf{E}(\mathbf{r}'),
\end{equation}
where $\mathbf{E}_0$ is the applied field, and $G_{1/d}$ is the $d$-dimensional dipolar Green function ($\theta$ is the Heaviside function, used to implement a principal value prescription at the origin)\cite{VANB61,*CHEW90}
\begin{equation}
\label{green}\mathsf{G}_{1/d}(\mathbf{r})=-\frac{\mathsf{
I}}{d}\delta(\mathbf{r})-\lim_{\eta\to
0}\theta(r-\eta)\frac{1}{S_d}\frac{1}{r^d}(\mathsf{I}-d\mathbf{\hat
r}\mathbf{\hat r}).
\end{equation}
In operator notation where, e.g.,
$\varepsilon(\mathbf{r})$ is understood as the bi-variate operator
$\varepsilon(\mathbf{r}|\mathbf{r}')\equiv \varepsilon(\mathbf{
r})\delta^{(d)}(\mathbf{r}-\mathbf{r}')$ so that $(\varepsilon
\mathbf{E})(\mathbf{r})=\int \mathrm{d}^d\!r\,\varepsilon(\mathbf{r}|\mathbf{r}')\mathbf{E}(\mathbf{r}')$,
Eq.\ (\ref{inte}) reads:
\begin{equation}
\label{eq:lsop}
E=E_b+G_{1/d}(\delta\varepsilon/\varepsilon_b) E.
\end{equation}

From this point on, we depart from the usual treatment. A modified Green operator $G_\ell$ is considered, in which the singularity at the origin is ``renormalized'' by introducing  \cite{PELL93} an arbitrary ``inner'' depolarization parameter $\ell$, of a more fundamental nature than $\ell_e$. By definition,
\begin{equation}
\label{gelldef}
\mathsf{G}_{1/d}(\mathbf{r})\equiv
-\ell\,\mathsf{I}\,\delta(\mathbf{r})+\mathsf{G}_\ell(\mathbf{r}).
\end{equation}
Writing the prefactor of the local (Dirac) part of $G_\ell$ as
\begin{equation}
\Delta\equiv \ell-1/d,
\end{equation}
one has by Eq.\ (\ref{green}) and definition (\ref{gelldef}),
\begin{equation}
\label{greenl}
\mathsf{G}_\ell(\mathbf{r})=\Delta\mathsf{
I}\delta(\mathbf{r})-\lim_{\eta\to
0}\theta(r-\eta)\frac{1}{S_d}\frac{1}{r^d}(\mathsf{I}-d\mathbf{\hat
r}\mathbf{\hat r}).
\end{equation}
Introducing moreover a ``screened'' electric field
\begin{equation}
\mathbf{E}_\ell\equiv
\left[1+\ell\frac{\delta\varepsilon(\mathbf{r})}{\varepsilon_b}\right]\mathbf{E},
\end{equation}
and a modified permittivity contrast
\begin{eqnarray}
\label{ul}
u_\ell(\mathbf{r})&\equiv&\frac{
[\delta\varepsilon(\mathbf{r})/\varepsilon_b]}{
1+\ell[\delta\varepsilon(\mathbf{r})/\varepsilon_b]}
=\frac{\varepsilon(\mathbf{r})-\varepsilon_b}
{\ell\varepsilon(\mathbf{r})+(1-\ell)\varepsilon_b},
\end{eqnarray}
Eq.\ (\ref{inte}) in transformed into the equivalent equation \cite{PELL93}
\begin{equation}
\label{intel} E_\ell=E_0+G_\ell\,u_\ell\,E_\ell.
\end{equation}

For $\ell=0$, $u_0(\mathbf{r})$ is the scaled dielectric contrast. For $\ell=1/d$, $u_{1/d}$ is (up to a prefactor $v\varepsilon_b$)
the dielectric polarizability of a spherical cell, and $\mathbf{E}_{1/d}(\mathbf{r})$ is the local (Lorentz) field impinging on it \cite{BROW55}. For $\ell\not=1/d$, some screening effects are implemented at the level of the cell. This possibility of splitting the Green function has been noticed previously \cite{MILT90} to the purpose of improving convergence in solving Eq.\ (\ref{inte}) by iterations (see also \cite{MILT02} p.\ 302). Although related, our aim is somewhat different.

\subsection{Multiple-scattering expansions}
\label{sec:msinterp}
Both  (\ref{eq:lsop}) and (\ref{intel}) are continuum analogues of scattering equations for finite-size
scatterers. It proves convenient to emphasize the connection by casting the continuum problem into a multiple-scattering framework. The \emph{scattering potential}
$\mathsf{u}_{\ell\mathbf{y}}$ of the material element at
$\mathbf{y}$ is introduced as
\begin{equation}
\label{uver}
\mathsf{u}_{\ell\mathbf{y}}(\mathbf{r}|\mathbf{r}')\equiv v\,
u_\ell(\mathbf{y})\delta(\mathbf{r}-\mathbf{y})
\delta(\mathbf{r}'-\mathbf{y})\,\mathsf{I}.
\end{equation}
Setting $\mathsf{G}_\ell(\mathbf{r}|\mathbf{r}')\equiv\mathsf{
G}_\ell(\mathbf{ r}-\mathbf{ r}')$ Equ.\ (\ref{intel}) is rewritten as
\begin{equation}
\mathbf{E}_\ell(\mathbf{ r})=\mathbf{ E}_0(\mathbf{ r})+\int
\frac{\mathrm{d\!}^d\!y}{v} \mathrm{d\!}^d\!r_2 \mathrm{d\!}^d\!r_1\,\mathsf{ G}_\ell(\mathbf{
r}|\mathbf{ r}_2)\mathsf{ u}_{\ell\mathbf{ y}}(\mathbf{
r}_2|\mathbf{ r}_1)\mathbf{ E}_\ell(\mathbf{ r}_1).
\end{equation}
In order to distinguish between integrations over `in' and `out' space
variables $\mathbf{r}_{1,2}$ and summations over scattering cells, let
$u_{\ell\mathbf{y}_j}\equiv u_{\ell j}$ and make the formal
replacement $\int (\mathrm{d}^d\!y/v)\to \sum_j$. Then (\ref{intel})
takes the form
\begin{equation}
\label{intel2} E_\ell=E_0+G_\ell\sum_j u_{\ell
j}E_\ell.
\end{equation}
Define now the Green function $G$ associated to $E_\ell$
by the equation $E_\ell=G G_\ell^{-1}E_0$, where
$G_\ell^{-1}E_0$ is the source. Then
\begin{equation}
\label{g} G=G_\ell+G_\ell\sum_j u_{\ell j}G.
\end{equation}

The individual transition operator, also called $T$-matrix (see \cite{ZAKH12} for a comprehensive list of references), completely characterizes the polarizability properties of the $i^{\text{th}}$ scatterer. It reads
\begin{equation}
\label{ti}
t_i\equiv u_{\ell i}(1-G_\ell u_{\ell i})^{-1}.
\end{equation}
Introducing  the Green function for the {\em local field} impinging
on the scatterer, namely, $G_i\equiv (1-G_\ell u_{\ell i})G$; the
polarization of the scatterer $P_i\equiv t_i G_i$; and finally its
``dressed'' $T$-matrix, $\widetilde{T}_i\equiv P_i G_\ell^{-1}$, which accounts for
corrections due to the other scatterers, one obtains from (\ref{g})
the familiar multiple-scattering equations
\cite{WATE61}:
\begin{subequations}
\label{difmul}
\begin{eqnarray}
\label{difmul1} G&=&G_\ell+G_\ell \mathcal{T}
G_\ell,\\
\label{difmul2}
\mathcal{T}&\equiv&\sum_i \widetilde{T}_i,\quad
\widetilde{T}_i=t_i+t_i G_\ell \sum_{j\neq i}
\widetilde{T}_j,
\end{eqnarray}
\end{subequations}
where $\mathcal{T}$ is the $T$-matrix of the whole system. We observe that since scattering events occur in succession between different scatterers, the Dirac term of $G_\ell$ is irrelevant in $\mathcal{T}$. Thus, we replace $G_\ell$ by $G_{1/d}$ in (\ref{difmul2}) to indicate that this propagator does not depend on $\ell$.

Next, a $n$-body (``virial") expansion of
the form ${\cal T}=\sum_{n\geq 1} {\cal
T}^{(n)}$ is written down, where the $n$-body term
\begin{equation}
\label{groupexp} {\cal T}^{(n)}=\sum{}'\,
T^{(n)}_{(i_1,i_2,\cdots,i_n)}
\end{equation}
is a sum over all possible scattering sequences on $n$ distinct scatterers.
The prime indicates that the sum is carried out first over label $i_1$, then
over $i_2\neq i_1$, etc., then over $i_n\not\in\{i_1,i_2,\cdots,i_{n-1}\}$, so that
$T^{(n)}_{(i_1,i_2,\ldots,i_n)}$ depends on label ordering.  This question was solved
long ago by Peterson and Str\"om \cite{PETE73}. We provide in Appendix \ref{appA}
another different approach to their result. Specifically, we show that:
\begin{subequations}
\label{nord}
\begin{eqnarray}
\label{nord1}
T^{(n)}_{(i_1,i_2,\ldots,i_n)}&=&T^{(n-1)}_{(i_1,i_2,\ldots,i_{n-1})}G_{1/d}
t_{i_n}\nonumber\\
&&\times\left(1-G_{1/d} S_{i_1\ldots i_{n-1}}G_{1/d} t_{i_n}\right)^{-1}\nonumber\\
&&\times\left(1+G_{1/d} S_{i_1\ldots i_{n-1}}\right),\\
\label{nord2} S_{i_1\ldots i_n}&=&G_{1/d}^{-1}\left[\left(1+G_{1/d}
t_{i_n}\right)\right.\nonumber\\
&&\times\left(1-G_{1/d} S_{i_1\ldots i_{n-1}} G_{1/d}
t_{i_n}\right)^{-1}\nonumber\\
&&\times\left.\left(1+G_{1/d} S_{i_1\ldots i_{n-1}}\right)-1\right],\\
\label{tone} T^{(1)}_i&=&S_i\equiv t_i.
\end{eqnarray}
\end{subequations}
In particular, the two-body term, to be used hereafter, is \cite{PETE73,NIEU87}:
\begin{eqnarray}
&&\hspace{-1em}T^{(2)}_{(i_1,i_2)}\!\!=\!\!t_{i_1}G_{1/d}t_{i_2}
\left(1\!-\!G_{1/d}t_{i_1}G_{1/d}t_{i_2}\right)^{-\!1}\!\!(1\!+\!G_{1/d}t_{i_1})\nonumber\\
&&{}=t_{i_1}G_{1/d}t_{i_2}\nonumber\\
&&+t_{i_1}G_{1/d}t_{i_2}G_{1/d}t_{i_1}G_{1/d}t_{i_2}
\left(1-G_{1/d}t_{i_1}G_{1/d}t_{i_2}\right)^{-1}\nonumber\\
\label{ttbt}
&&{}+t_{i_1}G_{1/d}t_{i_2}G_{1/d}t_{i_1}
\left(1-G_{1/d}t_{i_2}G_{1/d}t_{i_1}\right)^{-1}.
\end{eqnarray}
In the last writing, the term $t_{i_1}G_{1/d}t_{i_2}$ has been singled out for convenience, in the perspective of using Eq.\ (\ref{eq:twobt}) below. Although we disregard it for simplicity, the three-body term \cite{PETE73} has been considered by Cichocki and Felderhof \cite{CICH89b}.

\subsection{Ensemble averages, self-energy and effective permittivity}
\label{sec:ensav}
To address substitutional or positional randomness, ensemble averages over disorder, denoted
by $\langle\cdot\rangle$, are carried out \cite{FRIS68}. Due to statistical homogeneity, averaged operators are
translation-invariant. Introducing the complete scattering potential $U_\ell\equiv\sum_i
u_{\ell i}$ associated to the whole set of heterogeneities, the so-called self-energy (or coherent potential) operator $\Sigma_\ell$ of the averaged Green function $\langle G\rangle$ \cite{ROWL09} is defined by $\langle U_\ell
G\rangle\equiv \Sigma_\ell\langle G\rangle$, and from (\ref{g}) follows the Dyson equation \cite{DYSO49}
\begin{equation}
\label{avg} \langle G\rangle=(G_\ell^{-1}-\Sigma_\ell)^{-1}.
\end{equation}
The kernel $\mathsf{\Sigma}_\ell(\bf{r}|\bf{r}')=\mathsf{\Sigma}_\ell(\bf{r}-\bf{r}')$ possesses one local and one non-local part. It can be written as the Green function (\ref{green}) in the form
\begin{equation}
\label{eq:sigdef}
\mathsf{\Sigma}_\ell(\mathbf{r})=\Sigma_\ell^{\rm loc}\mathsf{I}\delta(\mathbf{r})+\mathsf{\Sigma}^{\rm nloc}_\ell(\mathbf{r}),
\end{equation}
where $\Sigma_\ell^{\rm loc}$ is a scalar, and where the same principal-value prescription as in (\ref{green}) applies to the non-local term. The necessity of distinguishing between the local and non-local part will show up when comparing theory to our reference data.

Similarly, the effective permittivity $\varepsilon_e$ is a non-local operator \cite{BARR07,*NIEZ10}, of the same generic form, and is defined through the equality $\langle \varepsilon \mathbf{
E}\rangle=\varepsilon_e\langle E\rangle$; that is,
\begin{equation}
\langle\varepsilon(\mathbf{ r})\mathbf{E}(\mathbf{r})\rangle=\int
\mathrm{d}^d\!r'\, \varepsilon_e(\mathbf{r}-\mathbf{
r}')\langle\mathbf{E}(\mathbf{r}')\rangle.
\end{equation}
Definitions of $E_\ell$, $\varepsilon_e$ and
$\Sigma_\ell$ lead to \cite{PELL93}
\begin{equation}
\label{epef}
\varepsilon_e=\varepsilon_b[1+(1-\ell)\Sigma_\ell][1-\ell
\Sigma_\ell]^{-1}.
\end{equation}
This formal equation becomes algebraic when Fourier transforms of the kernels are used. In particular, it holds for the volume integrals of the kernels. We recall that the effective $\ell_e(f)$ is obtained from $\varepsilon_e$ by means of (\ref{eq:elleps}).

The configurational average of (\ref{difmul1}) yields
\begin{equation}
\langle G\rangle=G_\ell +G_\ell\langle\mathcal{T}\rangle G_\ell.
\end{equation}
Comparing this equation with (\ref{avg}) provides the relationship:
\begin{equation}
\label{set} \Sigma_\ell=\langle {\cal T}\rangle[1+G_\ell\langle
{\cal T}\rangle]^{-1}.
\end{equation}
For an infinite system, this equation is merely formal and only has a meaning as a perturbative series, because $\langle\mathcal{T}\rangle$ involves conditionally-convergent integrals. Expanding (\ref{set}), and using (\ref{groupexp}) and (\ref{nord}), gives $\Sigma_\ell$ in the form of a $n$-body expansion, where one- and two-body contributions are read from (\ref{tone}) and (\ref{ttbt}):
\begin{subequations}
\label{eq:nbt}
\begin{eqnarray}
\label{eq:onebt}
&&\Sigma_\ell=\sum_{n\geq 1}\Sigma_\ell^{(n)},\quad\text{where}\quad
\Sigma_\ell^{(1)}=\sum_{i_1}\langle
t_{i_1}\rangle;\\
\label{eq:twobt}
&&\Sigma_\ell^{(2)}=
\sum_{i_1,i_2\not=i_1}\hspace{-1ex}\langle
T^{(2)}_{(i_1,i_2)}\rangle-\hspace{-1ex}\sum_{i_1,i_2}\langle
t_{i_1}\rangle G_\ell \langle t_{i_2}\rangle;\,\,\text{etc.}\qquad
\end{eqnarray}
\end{subequations}
The long-range part of the second term of $\Sigma_\ell^{(2)}$ cancels out the first term of the average of (\ref{ttbt}). The remaining terms involve only absolutely-convergent integrals, with an integrand decaying at least as $G_{1/d}^2\sim r^{-2d}$ as $r\to\infty$. This property holds at each order (the perturbative expansion of $\Sigma$ has a special -- ordered -- type of cumulant structure \cite{TERW74,MICH89}), which implies that $\Sigma$ is independent of the sample shape in the infinite-volume limit \cite{FELD82a,MICH89}.

\subsection{Parameters $s$ and $q$, and clustering}
\label{sec:params}
Before computing the self-energy, we remove some indeterminacies of the continuum theory. The latter admits two natural adjustable parameters. A first parameter $s>0$ stems from remarking that $\eta$ in Eq.\ (\ref{green}) must be of order $a$, which we write $\eta=s a$. We interpret $s/2$ as a measure of an \emph{effective radius} of neighboring inclusions, to be further constrained in Sec.\ \ref{sec:gmgt}. The value $s=2$ ($\eta=2 a$) corresponds to setting a hard-sphere-type exclusion distance between polarizable point inclusions. However, we allow here for smaller or larger values to tune the strength of dipolar interactions, so as to compensate for the lack of explicit higher-order multipoles in interactions between finite-size inclusions \cite{JEFF73,FELD82d}.

To work out the above equations, we need an expression of $t_i$, defined formally by Eq.\ (\ref{ti}). Dropping the index $i$ for brevity, and expanding, one has $t=u_{\ell}\sum_{k\geq 0}(G_\ell u_{\ell})^k$. Powers are evaluated by means of definitions (\ref{greenl}) and (\ref{uver}), but this involves squares of Dirac distributions: consider for instance $u_{\ell}(G_\ell u_{\ell})^2$, which reads
\begin{eqnarray}
&&\int \smash{\prod_{k=1}^4} d^d\!r_k\,u_{\ell\mathbf{y}}(\mathbf{r}|\mathbf{r}_1)
G_\ell(\mathbf{r}_1-\mathbf{r}_2)u_{\ell\mathbf{y}}(\mathbf{r}_2|\mathbf{r}_3)\times\nonumber\\
&&\hspace{2.5cm}{}\times G_\ell(\mathbf{r}_3-\mathbf{r}_4)u_{\ell\mathbf{y}}(\mathbf{r}_4|\mathbf{r}')\nonumber\\
&=&v^3 u_\ell(\mathbf{y})^3\mathsf{G}_\ell(\mathbf{0})^2\delta(\mathbf{r}-\mathbf{y})\delta(\mathbf{r}'-\mathbf{y})\nonumber\\
&=&v\, u_\ell(\mathbf{y})^3\Delta^2[v\delta(\mathbf{0})]^2\delta(\mathbf{r}-\mathbf{y})\delta(\mathbf{r}'-\mathbf{y}).
\end{eqnarray}
As discussed in Appendix \ref{app:distmult}, we use the prescription
\begin{equation}
\label{eq:delta2}
v\delta(\mathbf{r})^2\equiv q\delta(\mathbf{r}).
\end{equation}
The number $q>0$, a mathematical and physical necessary addition when $\Delta\not=0$, is the second parameter of the theory. We can then write $v\delta(\mathbf{0})=q$. The ``infinitely large'' number $v^{-1}$ gives the physical ``order of magnitude of $\delta(\mathbf{0})$'' in this problem. Thus,
\begin{equation}
\label{tiell}
t=u_{\ell}\sum_{k\geq 0}(q\Delta)^k u_{\ell}^k=u_\ell(1-q\Delta u_\ell)^{-1}=u_{\widetilde{\ell}},
\end{equation}
where
\begin{equation}
\label{eq:ltdef}
\widetilde{\ell}\equiv\ell-q\Delta=(1-q)\ell+q(1/d);
\end{equation}
that is, with $t(\mathbf{y})\equiv u_{\widetilde{\ell}}(\mathbf{y})$
\begin{equation}
\label{eq:taggr}
\mathsf{t}_\mathbf{ y}(\mathbf{ r}|\mathbf{ r}')=v\,t(\mathbf{ y})\delta(\mathbf{ r}-\mathbf{ y})\delta(\mathbf{
r}'-\mathbf{ y})\,\mathsf{I}.
\end{equation}
We assume that $0\leq q\leq 1$, so that $\widetilde{\ell}$ can be interpreted as a weighted average of $\ell$ and $1/d$.

\begin{figure}[!ht]
\centering
\includegraphics[width=7.0cm]{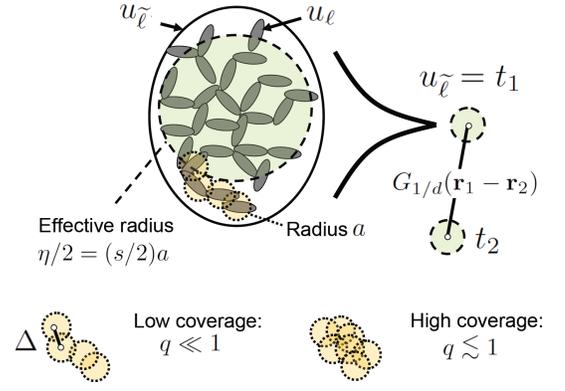}
\caption{\label{fig:fig2} (Color online) Clustering interpretation of the $s$ and $q$ parameters.}
\end{figure}
We propose the following interpretation of $q$, illustrated by Fig.\ \ref{fig:fig2}, where ellipsoidal shapes are meant to indicate that inclusions have an effective depolarization factor different from $1/d$ (the orientation of the ellipsoids in the drawing is irrelevant). Each scatterer $t_i$ is viewed as a aggregate of screened polarizable elements of polarizability proportional to $u_\ell$, \emph{of same permittivity}, embedded in medium $\varepsilon_b$. Their degree of clustering is adjusted through $q$. We interpret the latter as a coverage/spreading parameter for aggregated elements. In the figure, the small bar associated to $\Delta$ represents some spreading of $\delta(\mathbf{r})$ (see Appendix \ref{app:distmult}). When $q=0$, $t=u_\ell$ and the elements are considered separately; when $q=1$ they gather as a compact isotropic (spherical) inclusion of polarizability $t=u_{\frac{1}{d}}$. Setting $\ell\equiv 1/d$ suppresses the influence of $q$: the aggregate reduces to one single spherical inclusion in this case also. The aggregate is represented as a whole in a mean-field way in the sense that electrostatic interactions between its components are approximated by the local part $\Delta$ of $G_\ell$. Moreover, aggregates are treated as point-like polarizable objects [Eq.\ (\ref{eq:taggr})] when it comes to considering their mutual dipolar interactions via $G_\ell$ or $G_{1/d}$ (of which only the non-local part is relevant here; see Sec.\ \ref{sec:msinterp}).

To summarize, introducing $0<q<1$ makes the above theory a simplified one of interacting aggregates, in which the difference between $\ell$  and $\widetilde{\ell}$ distinguishes between inclusions and aggregates thereof.
\subsection{Self-energy to two-body order}
\label{sec:tba}
The self-energy follows from (\ref{eq:nbt}). Considering only volume-integrated kernels (with a slight abuse of notation),
\begin{eqnarray}
\Sigma^{(1)}&=&\Sigma^{(1){\rm loc}}=\int\mathrm{d}^{d}r\Sigma^{(1)}(\mathbf{r}-\mathbf{r}')
=\int\mathrm{d}^{d}\!r\sum_i\langle t_i(\mathbf{r}|\mathbf{r'})\rangle\nonumber\\
\label{eq:bigsig1}
&=&\int\mathrm{d}^{d}\!r\,\mathrm{d}^{d}\!y\,
\delta(\mathbf{r}-\mathbf{y})\delta(\mathbf{r}'-\mathbf{y})\langle t\rangle
=\langle t\rangle.
\end{eqnarray}
Likewise, from (\ref{ttbt}) and (\ref{eq:twobt}), $\Sigma^{(2)}=\Sigma^{(2){\rm loc}}+\Sigma^{(2){\rm nloc}}$ with
\begin{subequations}
\label{eq:sig2def}
\begin{eqnarray}
\label{eq:sig2defl}
&&\Sigma^{(2){\rm loc}}=-\Delta\langle t\rangle^2
+\int_{r>\eta}\hspace{-1.5em}\mathrm{d\!}^{d}r\left\langle\frac{v\,t^2\,\widetilde{t}\,\mathsf{G}_{1/d}^2(\mathbf{r})}{1-v^2 t\, \widetilde{t}\,\mathsf{G}_{1/d}^2(\mathbf{r})}\right\rangle,\\
\label{eq:sig2defnl}
&&\Sigma^{(2){\rm nloc}}=\int_{r>\eta}\hspace{-1.5em}\mathrm{d\!}^{d}r\left\langle\frac{v^2\,
t^2\,{\widetilde{t}}^2\,\mathsf{G}_{1/d}^3(\mathbf{r})}{1-v^2 t\,\widetilde{t}\,\mathsf{G}_{1/d}^2(\mathbf{r})}\right\rangle,
\end{eqnarray}
\end{subequations}
where $\eta=s a$, and  where $\mathbf{r}$ is the separation vector between two statistically uncorrelated cells of volume $v$. We denote their polarizabilities by $t$ and $\widetilde{t}$ to distinguish them. The configurational average over permittivities must be carried out independently on these quantities.

The first two terms of (\ref{ttbt}) are built on scattering sequences that start and end on the different scatterers $i_1\not =i_2$, and thus contribute to $\Sigma^{(2){\rm nloc}}$, which stands as a non-local susceptibility. Instead, the third one is made of closed scattering sequences that start and end on the same scatterer $i_1$, and so contributes to the local part $\Sigma^{(2){\rm loc}}$ as a renormalization (``dressing'') of the $T$-matrix of individual scatterers \cite{NIEU87}. Indeed, for $\Delta=0$, in the diagrammatic representation \cite{BERG81b,BARR89} where a line stands for a ``propagator'' $G_{1/d}$ and a $2n$-legged dot stands for the $n$th power of a $T$-matrix $t$,
\begin{subequations}
\begin{eqnarray}
\Sigma^{{(1)\rm loc}}&=&\diag{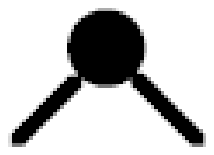}\\
\Sigma^{{(2)\rm loc}}&=&\diag{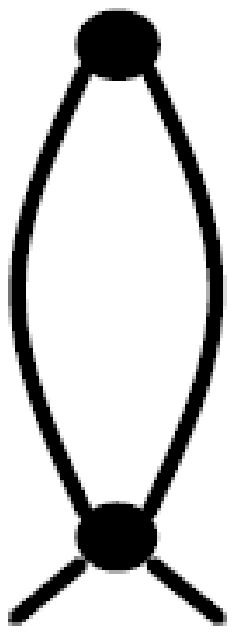}+\diag{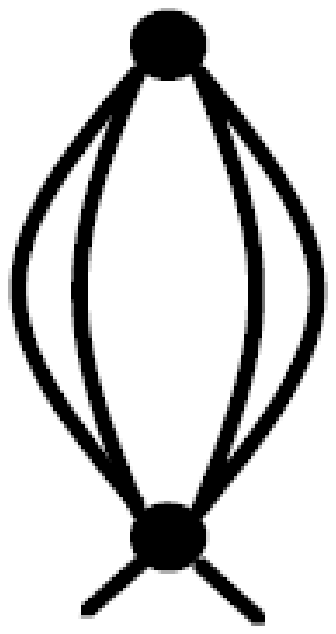}+\diag{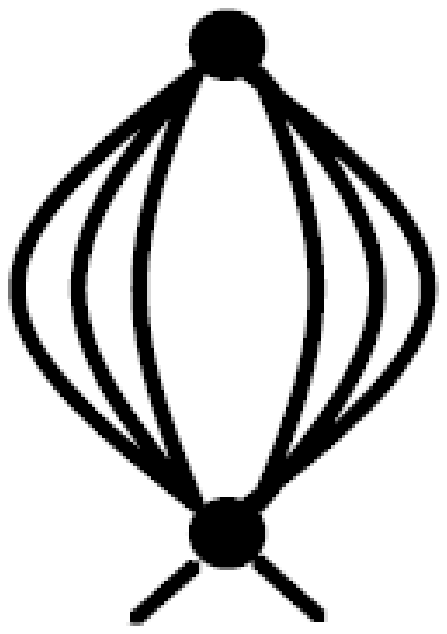}+\cdots\\
\Sigma^{{(2)\rm nloc}}&=&\raisebox{-1pt}{\diag{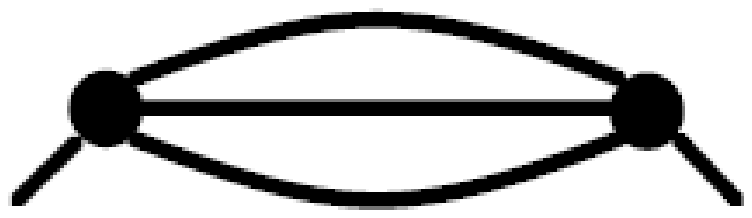}}+\raisebox{-3pt}{\diag{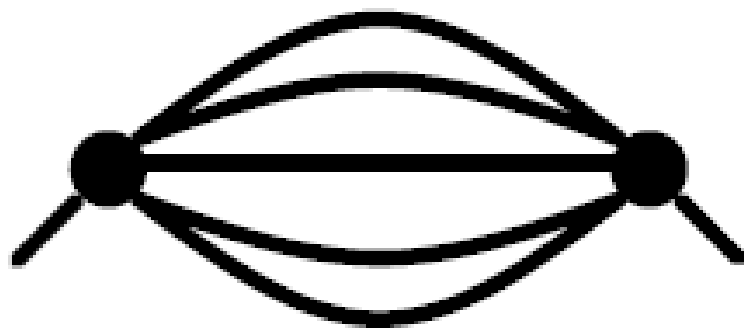}}+\raisebox{-5pt}{\diag{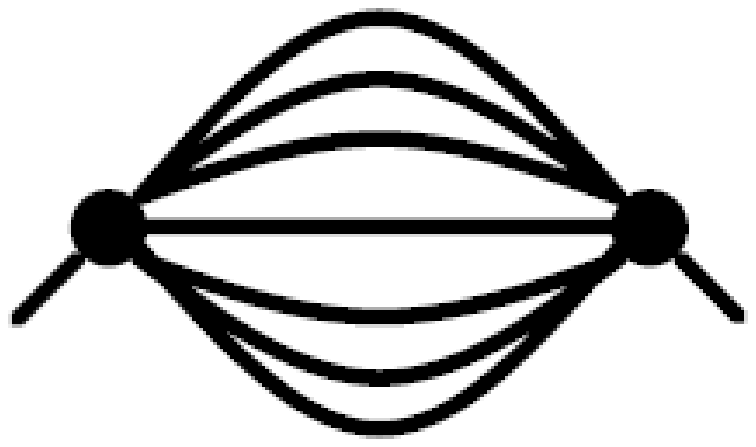}}+\cdots
\end{eqnarray}
\end{subequations}
As recalled in Sec.\ \ref{sec:ensav}, the absolute convergence of the integrals in $\Sigma^{(2)}$ results from the fact that the latter involves only terms with at least two propagators. When $\Delta\not=0$, the additional contribution $-\Delta\langle t\rangle^2$, which stems from the local part of $G_\ell$ in the last term of (\ref{eq:twobt}), is counted within $\Sigma^{{(2)\rm loc}}$ rather than $\Sigma^{{(2)\rm nloc}}$, in view of definition (\ref{eq:sigdef}) of the local and nonlocal parts of the $\Sigma$ operator.

In spite of the shorthand scalar notation employed in (\ref{eq:sig2def}) (all terms commute) matrix inverses are required. They follow from writing the nonlocal part of $\mathsf{G}_{1/d}$ as a linear combination of the projectors $\mathbf{\hat r}\mathbf{\hat r}$ and $\mathsf{I}-\mathbf{\hat r}\mathbf{\hat r}$. Carrying out the integrals over $\mathbf{r}$ in (\ref{eq:sig2def}) yields \cite{*[{}] [{. In this unpublished work, referred to in [11], two-body effective-medium theories for RRNs, and in the continuum, were considered. A s.c.\ condition $\Sigma_{1/d}=0$ was used. In the continuum, this is the theory of Sec.\ \ref{sec:gscd} with $\alpha=\beta=1$.}] PELL92}
\begin{subequations}
\label{eq:bigsig2}
\begin{eqnarray}
\label{sig2}
&&\Sigma^{(2){\rm loc}}=-\Delta\langle t\rangle^2
+d^{-2}(d-1)\left\langle t(t\widetilde{t})^{\frac{1}{2}}g(z)\right\rangle,\\
&&\Sigma^{(2){\rm nloc}}=d^{-2}(d-1)\left\langle t\widetilde{t}\,h(z)\right\rangle,\\
&&g(z)=\tanh^{-1}(z)+\tanh^{-1}\bigl((d-1)z\bigr),\\
&&h(z)=\frac{1}{2}\log\frac{1-z^2}{1-(d-1)^2z^2},\\
\label{eq:zdef}
&&z=\sqrt{t\widetilde{t}}/(d\, s^{d}).
\end{eqnarray}
\end{subequations}
We note for further use that, with principal determinations,
\begin{equation}
\label{eq:gph}
g(z)+h(z)=\log\frac{1+z}{1-(d-1)z}.
\end{equation}
Due to scale invariance, size $a$ is absent from these equations.

Gathering one- and two-body terms, the explicit expression of $\Sigma_\ell$ after having taken averages is as follows. We set $p_1=(1-f)$, $p_2=f$, and introduce $t_1$ and $t_2$ such that
\begin{equation}
t_{1,2}=\frac{\varepsilon_{1,2}-\varepsilon_b}{\widetilde{\ell}\varepsilon_{1,2}+(1-\widetilde{\ell})\varepsilon_b}.
\end{equation}
The quantities $t$ and $\widetilde{t}$ in Eq.\ (\ref{eq:sig2def}) take on values $t_1$ or $t_2$ with respective probabilities $p_1$ and $p_2$. Introduce moreover $z_{11}=t_1/(d s^d)$, $z_{12}=\sqrt{t_1 t_2}/(d s^d)$, $z_{22}=t_2/(d s^d)$. Then, from Eqs.\ (\ref{eq:bigsig1}) and (\ref{eq:bigsig2}), with $\langle t\rangle=p_1 t_1+p_2 t_2$,
\begin{subequations}
\label{eq:siglocnloc}
\begin{eqnarray}
\label{eq:sigelexpl}
&&\Sigma_\ell^{\rm loc}=\langle t\rangle-\Delta\langle t\rangle^2
+d^{-2}(d-1)\bigl[(p_1 t_1)^2 g(z_{11})\nonumber\\
{}+{}&&(p_2 t_2)^2 g(z_{22})+p_1 p_2(t_1+t_2)(t_1 t_2)^{\frac{1}{2}} g(z_{12})\bigr],\\
&&\Sigma_\ell^{nloc}=d^{-2}(d-1)\bigl[(p_1 t_1)^2h(z_{11})\nonumber\\
\label{eq:sigellnloc}
{}+{}&&(p_2 t_2)^2 h(z_{22})+2 (p_1 t_1)(p_2 t_2) h(z_{12})\bigr].
\end{eqnarray}
\end{subequations}
We remark that $h(z)\equiv 0$ for $d=2$, so that in two dimensions $\Sigma_\ell\equiv\Sigma_\ell^{\rm loc}$.

Two-body interactions terms in the effective permittivity of composites have been considered by numerous authors, many of who focused on interactions between spherical inclusions of finite size \cite{JEFF73,ALEX10}. Here, taking $d=3$, $\ell=1/3$, $s=2$, $\varepsilon_b=\varepsilon_1$, and $t_1=\alpha/(\varepsilon_1 v)$ where $\alpha$ is a polarizability, and $t_2=0$, and expanding $\varepsilon_e$ to order $O(f^2)$ returns a known expression of the two-body correction in the effective permittivity of a suspension of polarizable \emph{point} inclusions distributed according to the law of a hard-sphere gas \cite{BUCK55a,FELD82d}. An often-cited expression for this term \cite{PETE69} is only an approximate one.

\section{Effective-medium conditions}
\label{sec:emconds}
In this section, the general theory is completed by s.c.\ conditions, and exploited. Several schemes are possible. For clarity, a non-self-consistent setting is considered first.
\subsection{Theory of the Clausius-Mossoti type}
\label{sec:gmgt}
When used with the volume integrals of the kernels, Eq.\ (\ref{epef}) resembles the Clausius-Mossoti (CM) formula \cite{CLAU87,*MOSS50,*LORE80a,*LORE80b}, in which the polarizability of inclusions replaces the self-energy, and constitutes a generalization of the Maxwell--Garnett effective-medium formula \cite{MAXW04}. The latter holds for dilute systems of spherical inclusions embedded in a matrix. It is retrieved by letting $\ell=1/d$, fixing
the background permittivity to that of the matrix, and keeping
only in $\Sigma$ the one-body contribution (\ref{eq:onebt}).
Similar many-body generalizations of the CM formula by means of cluster
expansions have previously been worked out by Felderhof and
co-workers \cite{FELD82a,FELD83,CICH89a,CICH89b}, among others.

Whenever $\Sigma_\ell\not=0$ in Eq.\ (\ref{epef}), which produces CM-type estimates, a distinction must be made between the backgound ``inner'' permittivity  $\varepsilon_b$ and the overall one $\varepsilon_e$. It follows that the ``inner'' $\ell$ of the theory needs to be distinguished from the ``effective'' one, $\ell_e(f)$ obtained from $\varepsilon_e$ by means of Eq.\ (\ref{eq:elleps}).

In the rest of this Section and in the next one, we fix $\ell$ to its usual value $1/d$. This eliminates altogether the $q$ parameter, see Sec.\ \ref{sec:params}, and implies that $\Delta=0$ and
\begin{equation}
t_{1,2}=\frac{d(\varepsilon_{1,2}-\varepsilon_b)}{\varepsilon_{1,2}+(d-1)\varepsilon_b}.
\end{equation}

Dilute behavior in the CM-type approach is then as follows. Setting $\varepsilon_b=\varepsilon_1$ the $O(f)$ term of $\ell_e$ is readily obtained. Using identity (\ref{eq:gph}), the result reads
\begin{eqnarray}
\ell_e^{\rm CM}(f)&=&\frac{1}{d}+\frac{d-1}{d^2}\left[d\zeta_2-\log\frac{1+\zeta_2/s^d}{1-(d-1)\zeta_2/s^d}\right]f,\nonumber\\
\label{eq:zet2}
\zeta_2&\equiv&\frac{\varepsilon_2-\varepsilon_1}{\varepsilon_2+(d-1)\varepsilon_1}.
\end{eqnarray}
The medium being symmetric, the $O(f-1)$ behavior stems from interchanging $\varepsilon_1$ and $\varepsilon_2$, and from replacing $f$ by $1-f$. Letting $\ell_e(f)=1/d+\sigma_0 f+O(f^2)=1/d+\sigma_1(f-1)+O((f-1)^2)$, the dilute-limit slopes of the $\ell_e(f)$ graph in the infinite-contrast limit $\varepsilon_2\gg\varepsilon_1$ follow as
\begin{subequations}
\label{eq:sigmg}
\begin{eqnarray}
\label{eq:sig0mg}
\sigma_0^{\rm CM}&=&\frac{d-1}{d^2}\left[d-\log\frac{s^d+1}{s^d-(d-1)}\right],\\
\label{eq:sig1mg}
\sigma_1^{\rm CM}&=&\frac{d-1}{d^2}\left[\frac{d}{d-1}-\log\frac{s^d+1}{s^d-(d-1)^{-1}}\right].
\end{eqnarray}
\end{subequations}
By self-duality \cite{STRA77}, $\sigma_0^{\rm CM}=\sigma_1^{\rm CM}$ for $d=2$.

The cut of the logarithm in (\ref{eq:zet2}) materializes the two-body resonance spectrum \cite{CLER96} of the theory, and such resonances should not occur for positive $\varepsilon_{1,2}$; that is, $\ell_e$ must be real in statics. Assuming that $s$ does not depend on volume fractions, the requirement that the infinite-contrast limits of $\sigma_{0,1}$ be real implies the constraint
\begin{equation}
\label{eq:constrs}
s>s_{{\rm min}}\equiv(d-1)^{1/d}.
\end{equation}
This is a direct consequence of requiring that $(d-1)\max[|z_{11}|,|z_{12}|,|z_{22}|]<1$ in Eqs.\ (\ref{eq:siglocnloc}) to avoid the cuts  whatever $d$ and the contrast.

\begin{figure}[!ht]
\centering
\includegraphics[width=4.2cm]{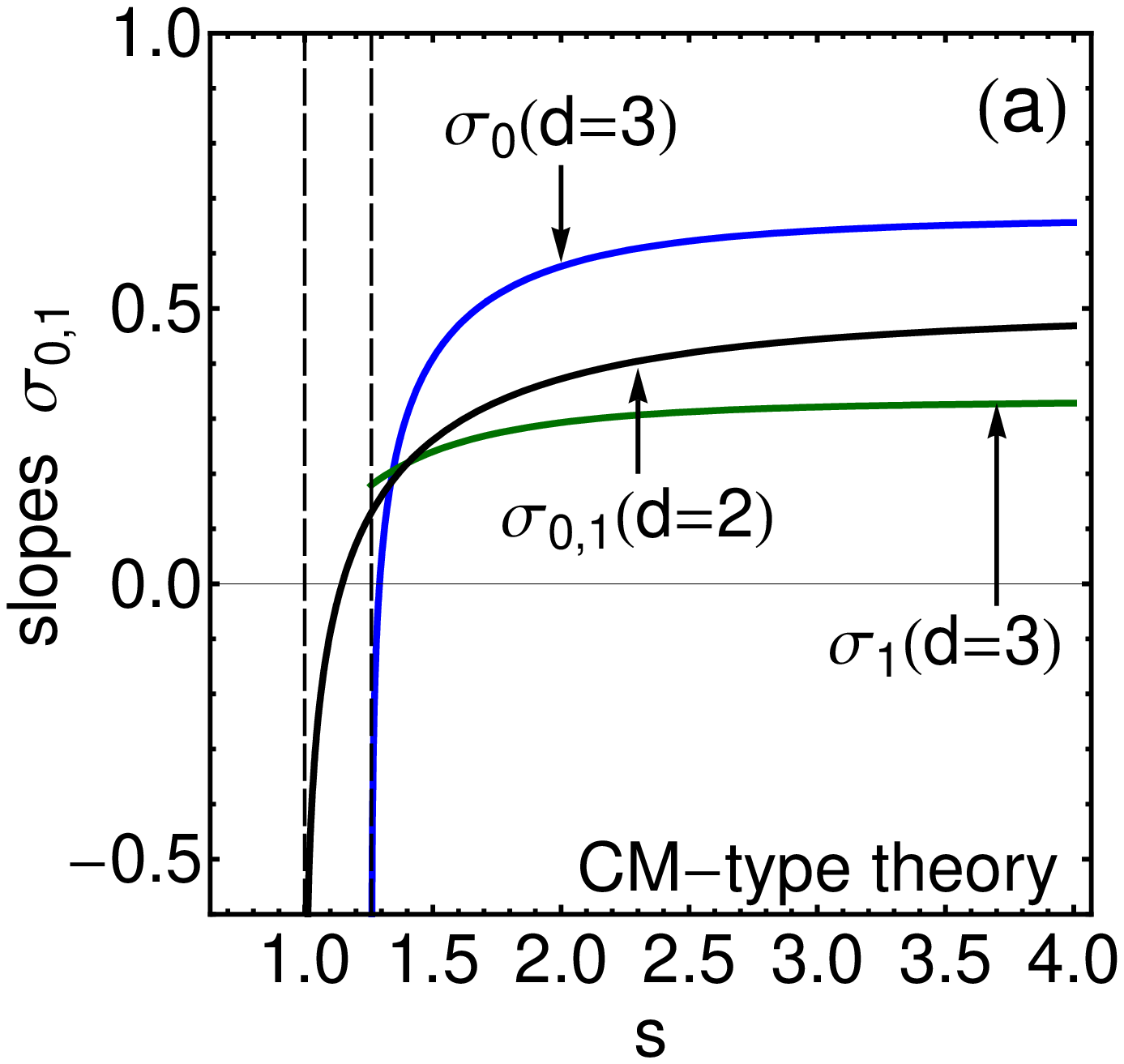}
\includegraphics[width=4.2cm]{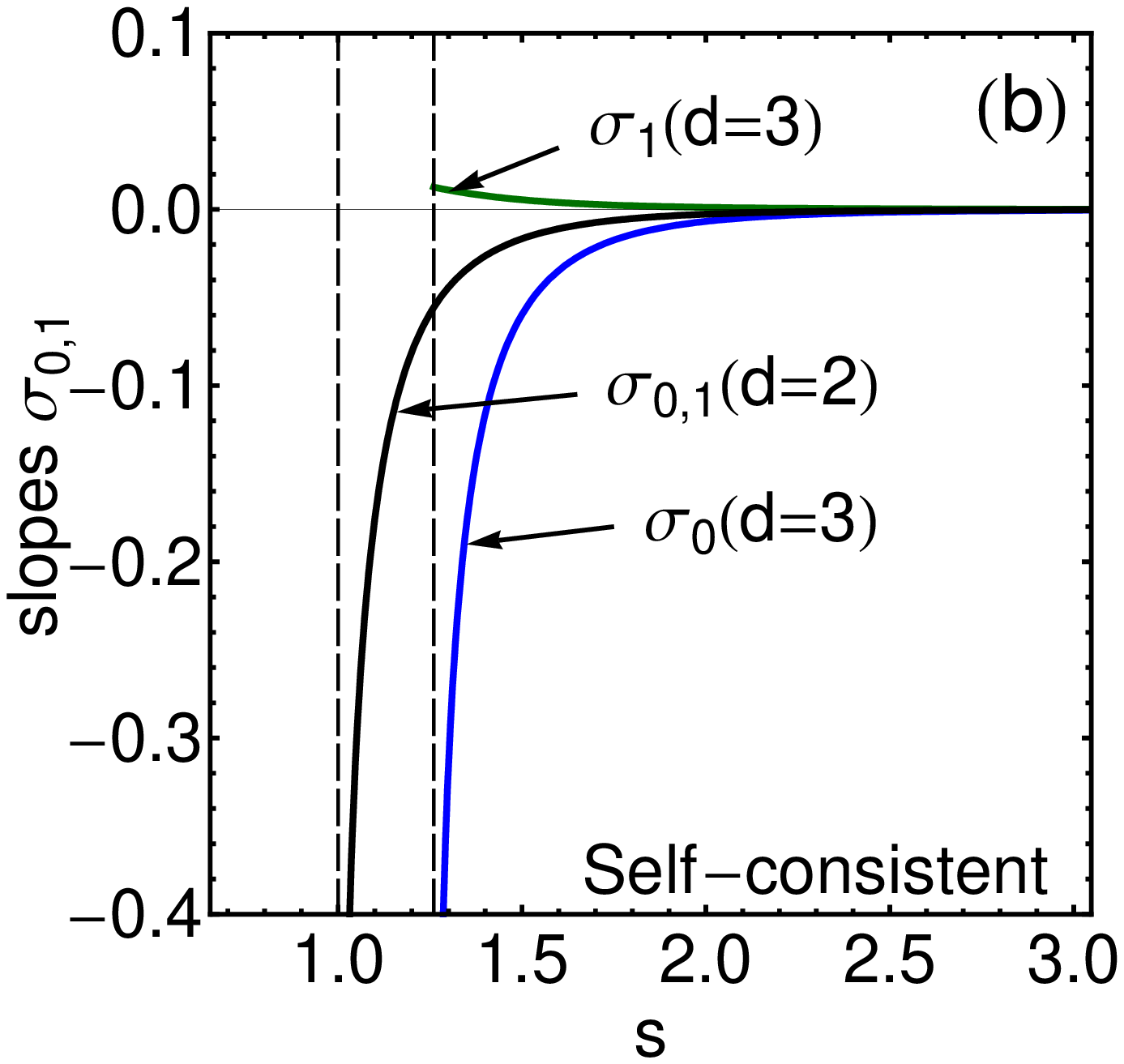}
\caption{\label{fig:fig3} (Color online) Slopes $\sigma_{0,1}$ in (a) the CM-type theory, and (b) generalized self-consistent theory, with $\ell=1/d$, vs. parameter $s$. Vertical dashed lines represent the $s_{\rm min}$ lower bounds.}
\end{figure}

Fig.\ \ref{fig:fig3}(a) represents the functions $\sigma_{0,1}^{\rm CM}(s)$. Slope $\sigma_0$ blows-up logarithmically near $s_{\rm min}$. Thus, for $d=2$, negative slopes $\sigma_{0,1}$ such as in Figs.\ \ref{fig:fig1}(a) and (c) are possible for $s$ close enough to $1$. For $d=3$, negative slopes $\sigma_0$ are obtained for $s$ close to $s_{\rm min}(d=3)\simeq 1.25992$, and the theory admits positive  $\sigma_1$ slopes: this is evocative of the PDA behavior of Fig.\ \ref{fig:fig1}(d) in the dilute limit, but the corresponding $\sigma_1$ values, of order $0.2$, outstrip those of simulations. Besides, this approach cannot reproduce the weak negative slope at $f=1$ of Fig.\ \ref{fig:fig1}(b).

\subsection{Generalized self-consistency with $\ell=1/d$.\\ Parameters $\alpha$ and $\beta$.}
\label{sec:gscd}
\begin{figure*}[t]
\centering
\includegraphics[width=4.3cm]{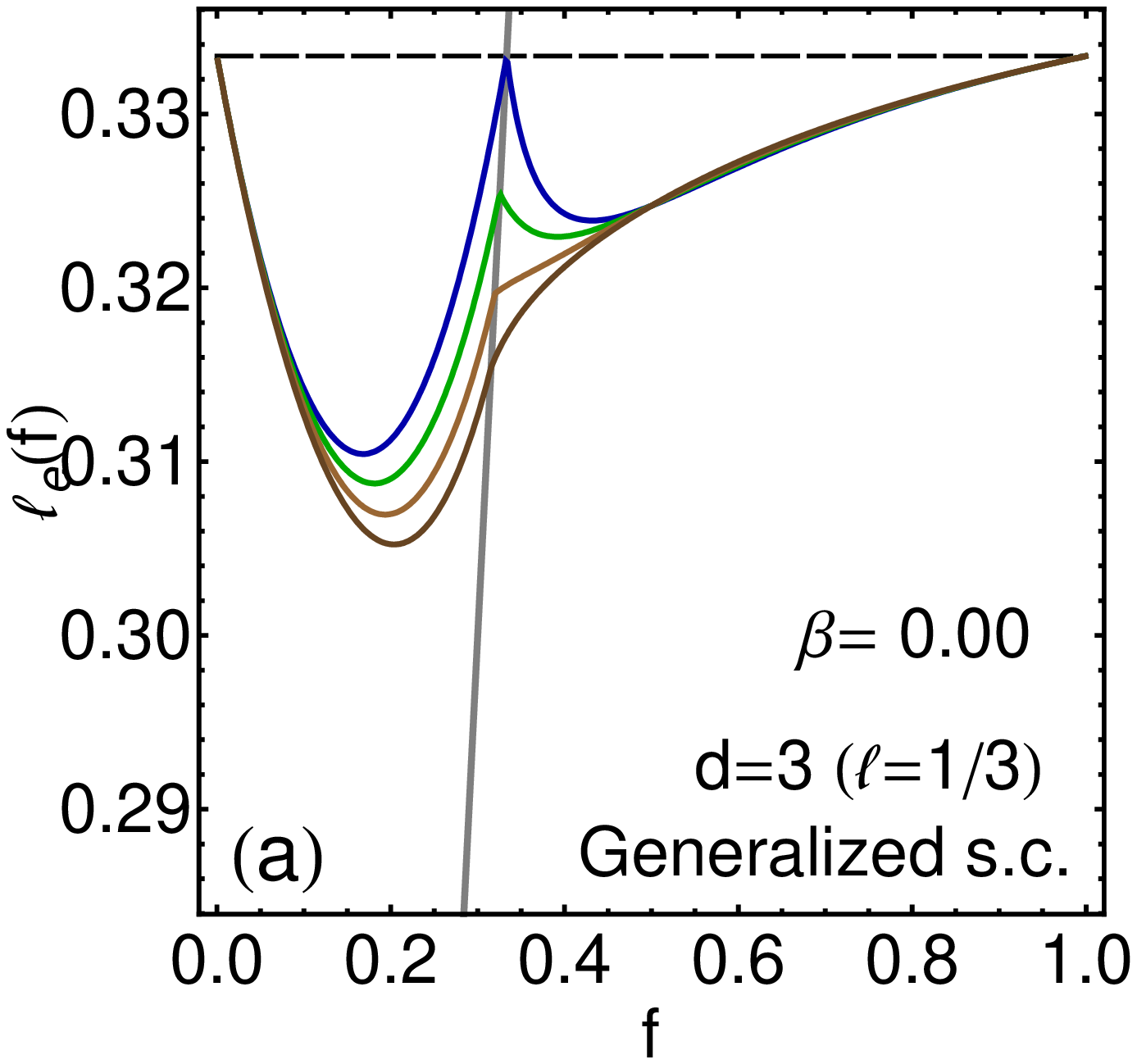}
\includegraphics[width=4.3cm]{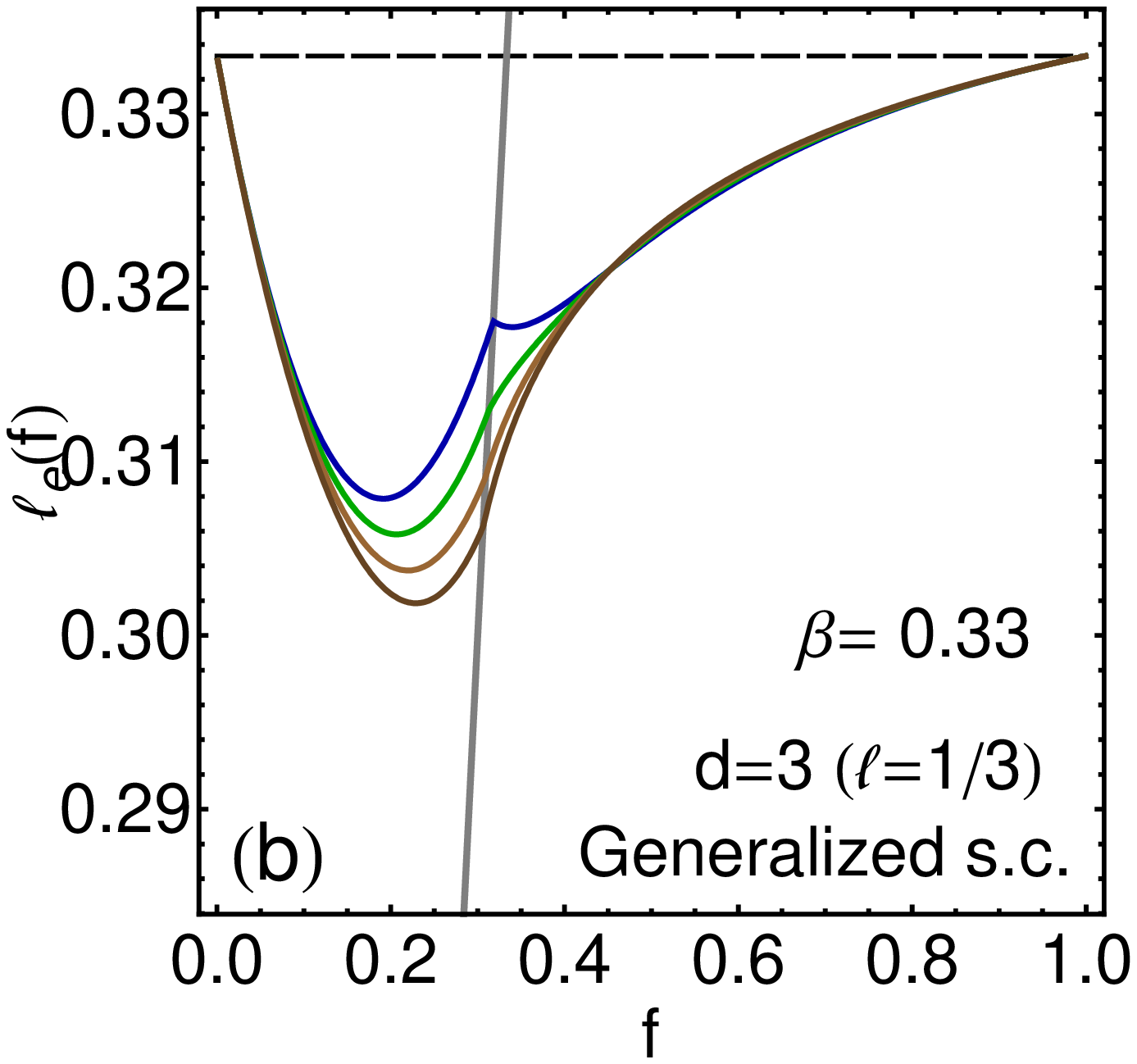}
\includegraphics[width=4.3cm]{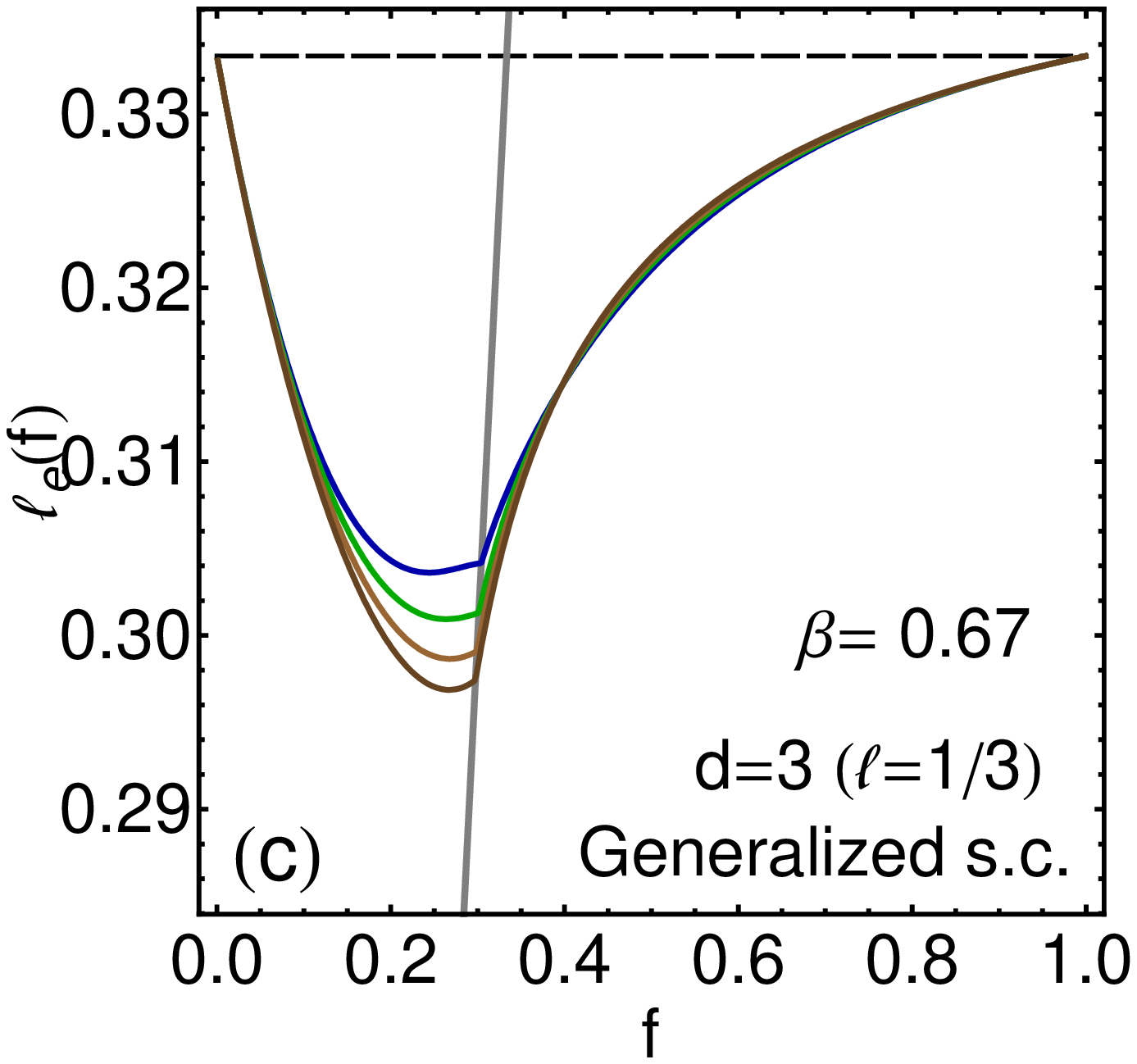}
\includegraphics[width=4.3cm]{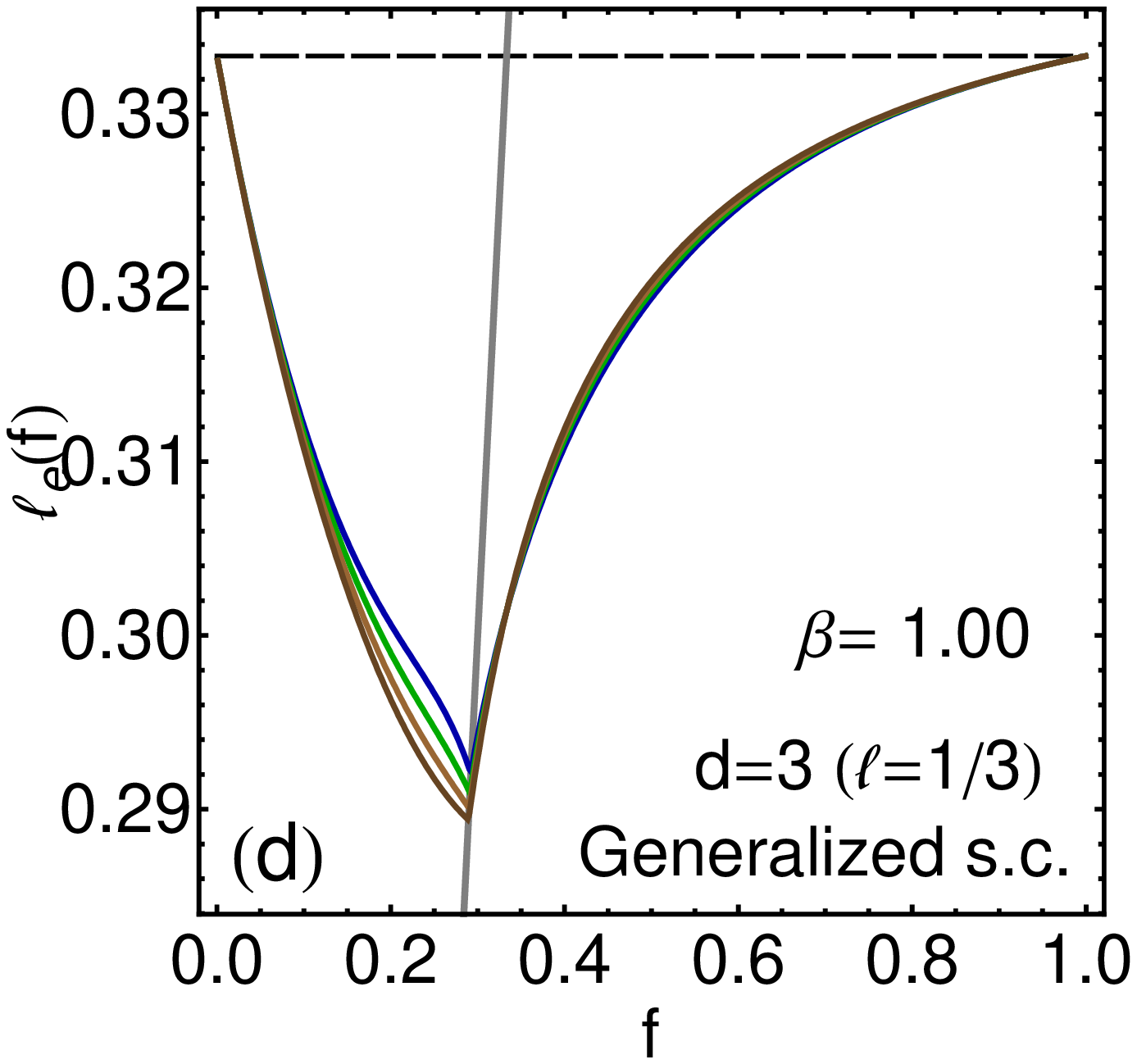}
\caption{\label{fig:fig4} (Color online) Dimension $d=3$. Dependence on $\beta$ (as indicated) of $\ell_e(f)$ in the generalized s.c.\ approach at infinite contrast, with $s=s_{\rm min}+0.05$ and $\alpha=0.00$, $0.33$, $0.67$ and $1.00$ in each figure (top to bottom).}
\end{figure*}
We keep using $\ell=1/d$. The exceeding high slopes (\ref{eq:sigmg}) are a consequence of their going to a finite limit as $s\to\infty$, because of the term $d\zeta_2$ within the brackets of Eq.\ (\ref{eq:zet2}). This term is traced to the one-body contribution $\Sigma^{(1)}$ to the self-energy, i.e., it already exists in the standard Maxwell-Garnett formula interpreted in terms of $\ell_e$. Powers of $\Sigma^{(1)}$ are generated to all orders of perturbations when expanding Eq.\ (\ref{epef}). The self-energy expansion (\ref{eq:nbt}) is essentially an asymptotic one, and the CM-like expression (\ref{epef}) does not perform well in reorganizing it into a physically-meaningful effective permittivity for all volume fractions.

A widely-used reshuffling device is to determine $\varepsilon_b$ self-consistently by the CPA condition $\Sigma\equiv\Sigma^{(1)}+\Sigma^{(2)}=0$, which implies that $\varepsilon_e=\varepsilon_b$. However this is not the only possibility: we can also consider the restricted one-body version $\Sigma^{(1)}=0$ \cite{SOVE67}, i.e., the BL condition; or even an intermediate ``local'' condition $\Sigma^{\rm loc}\equiv\Sigma^{(1)}+\Sigma^{(2)\rm loc}=0$, etc. To handle them all, and much more, we introduce new parameters $\alpha$ and $\beta$, and put forward the \emph{generalized s.c.\ condition}
\begin{equation}
\label{eq:gensc}
\Sigma^{(1)}+\alpha\Sigma^{(2){\rm loc}}+\beta\Sigma^{(2){\rm nloc}}=0\qquad 0\leq \alpha,\beta\leq 1.
\end{equation}
Parameter $\beta$ is irrelevant for $d=2$, according to our remark following Eq.\ (\ref{eq:sigellnloc}).
With the above condition, and setting $\delta\Sigma\equiv (1-\alpha)\Sigma^{(2){\rm loc}}+(1-\beta)\Sigma^{(2){\rm nloc}}$, Eq.\ (\ref{epef}) reduces to
\begin{equation}
\varepsilon_e=\varepsilon_b\frac{1+(1-1/d)\delta\Sigma}{1-\delta\Sigma/d},
\end{equation}
which, unless $\alpha=\beta=1$, remains of the CM type in spite of the s.c.\ condition.

As far as the $O(f)$ term in $\ell_e$ is concerned, the outcome of condition (\ref{eq:gensc}) is independent of $(\alpha,\beta)$; this property does not hold for $\ell$. With $\zeta_2$ as in (\ref{eq:zet2}), working out the dilute expansion indeed gives
\begin{equation}
\label{eq:lsc}
\ell_e^{\rm sc}(f)=\frac{1}{d}+\frac{d-1}{d^2}\left[d\frac{\zeta_2}{s^d}-\log\frac{1+\zeta_2/s^d}{1-(d-1)\zeta_2/s^d}\right]f.
\end{equation}
It is important to remark that with self-consistency, the slope at $f=0$ becomes a function of $\zeta_2/s^d$, which turns parameter $s$ into a polarizability rescaling factor. Infinite-contrast slopes follow as:
\begin{subequations}
\label{eq:sigsc}
\begin{eqnarray}
\label{eq:sig0sc}
\sigma_0^{\rm sc}&=&\frac{d-1}{d^2}\left[\frac{d}{s^d}-\log\frac{s^d+1}{s^d-(d-1)}\right],\\
\label{eq:sig1sc}
\sigma_1^{\rm sc}&=&\frac{d-1}{d^2}\left[\frac{d s^{-d}}{(d-1)}-\log\frac{s^d+1}{s^d-(d-1)^{-1}}\right].
\end{eqnarray}
\end{subequations}
They are drawn in Fig.\ \ref{fig:fig3}(b). They vanish in the limit $s\to\infty$ where two-body interactions are suppressed, so that the slopes are now a \emph{pure two-body effect}. However, whereas $\sigma_1$ takes on more realistic values, it still cannot be made negative for $d=3$.

Percolation occurs through $\varepsilon_b$. A second-degree polynomial equation for the percolation threshold $f_c$ is obtained as the infinite-contrast limit of (\ref{eq:gensc}), letting $\varepsilon_1=0$ first, then $\varepsilon_b=0$. For brevity, we no not reproduce its lengthy coefficients. The solution for $d=2$ is $f_c=1/2$ irrespective of the parameters. For $d=3$, the model percolates at the BL threshold $f_c=1/3$ if $\alpha=\beta=0$; otherwise, the leading-order term of the equation for $s$ near to $s_{\rm min}$ provides the asymptotic estimate
\begin{equation}
\label{eq:percoab}
f_c\sim\frac{1}{2}\left[\frac{6+\alpha\log 5-\beta \log(5/4)}{(\alpha+\beta)|\log(s-s_{\rm min})|}\right]^{1/2}.
\end{equation}
The lowest value of $f_c$ is obtained with $\alpha=\beta=1$, at fixed $s$, and $f_c$ can be made as small as needed by letting $s\to s_{\rm min}$.

Fig.\ \ref{fig:fig4} illustrates typical consequences on $\ell_e(f)$ of the generalized s.c.\ condition, in the three-dimensional case where variations are most conspicuous. As long as self-consistency involves two-body interactions, the threshold is lowered with respect to the BL value, with an intricate dependence on $\alpha$ and $\beta$. Their influence on $\ell_e$ is confined to a definite region around $f_c$, because the dilute-limit slopes do not depend on $\alpha$ and $\beta$. The effect of removing part of $\Sigma^{(2)\rm nloc}$ from the s.c.\ condition is interesting: in Fig.\ \ref{fig:fig4}(a), a graph shape with an upward cusp at $f_c$, akin to that of Fig.\ \ref{fig:fig1}(d) (highest contrast), is produced with $\beta=0$ and a small amount of $\Sigma^{(2)\rm loc}$ ($\alpha=0.33$); on the other hand, a downward cusp results from injecting a high amount of $\Sigma^{(2)\rm nloc}$ [$\beta=1$, Fig.\ \ref{fig:fig4}(d)].

\subsection{Generalized self-consistency with variable $\ell$}
\label{sec:gscl}
We can now discuss the effect of the ``inner'' depolarization variable $\ell$. Relaxing the condition $\ell=1/d$, the s.c.\ condition (\ref{eq:gensc}) for $\varepsilon_b$ still holds, but this time expressed in terms of $\Sigma_\ell$ with $\ell$ arbitrary. Since $\Delta=\ell-1/d\not=0$, the self-energy has one additional term in its ``loc'' part (\ref{eq:sigelexpl}), and the $q$ parameter introduced in Sec.\ \ref{sec:params} becomes operative. The theory of Sec.\ \ref{sec:gscd} is retrieved if $q\equiv 1$.  To determine $\ell$ we impose the supplementary s.c.\ condition,
\begin{equation}
\label{em1}
\langle u_\ell\rangle=0,
\end{equation}
which, in the interpretation of Sec.\ \ref{sec:params}, makes $\varepsilon_b$ the local effective medium surrounding the elementary screened elements within aggregates. This is Eq.\ (\ref{eq:brugellmod}), with $\varepsilon_e$ and $\ell_e$ replaced by $\varepsilon_b$ and $\ell$, respectively. Its solutions are
\begin{eqnarray}
\varepsilon_b^{\pm}&=&\frac{\varepsilon_1}{1-\ell}\left[\tau\pm\sqrt{\ell(1-\ell)\tau\chi+\tau^2}\right],\nonumber\\
\label{eq:ebsols}
\tau&=&\frac{1}{2}[(1-f-\ell)+(f-\ell)\chi],\quad \chi\equiv\varepsilon_2/\varepsilon_1.
\end{eqnarray}
Solutions $\ell$ follow from using these expressions in the generalized CPA condition (\ref{eq:gensc}). If $\alpha=1$ (case $d=2$ where $\beta$ is irrelevant), or $\alpha=\beta=1$ (case $d\geq 3$), then $\Sigma=0$ so that $\varepsilon_e=\varepsilon_b$ and $\ell_e=\ell$.
\begin{figure}[!ht]
\centering
\includegraphics[width=4.25cm]{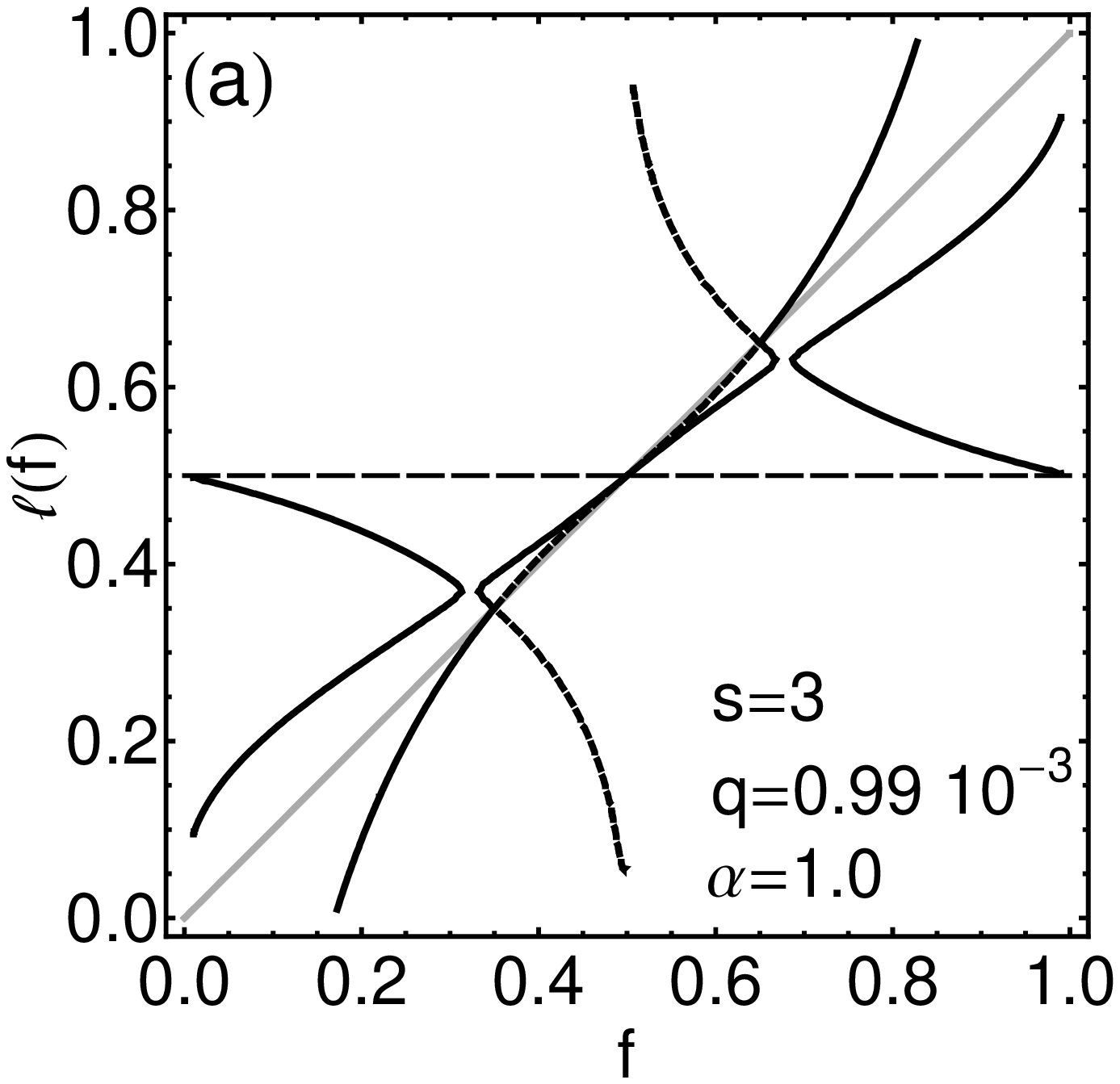}
\includegraphics[width=4.25cm]{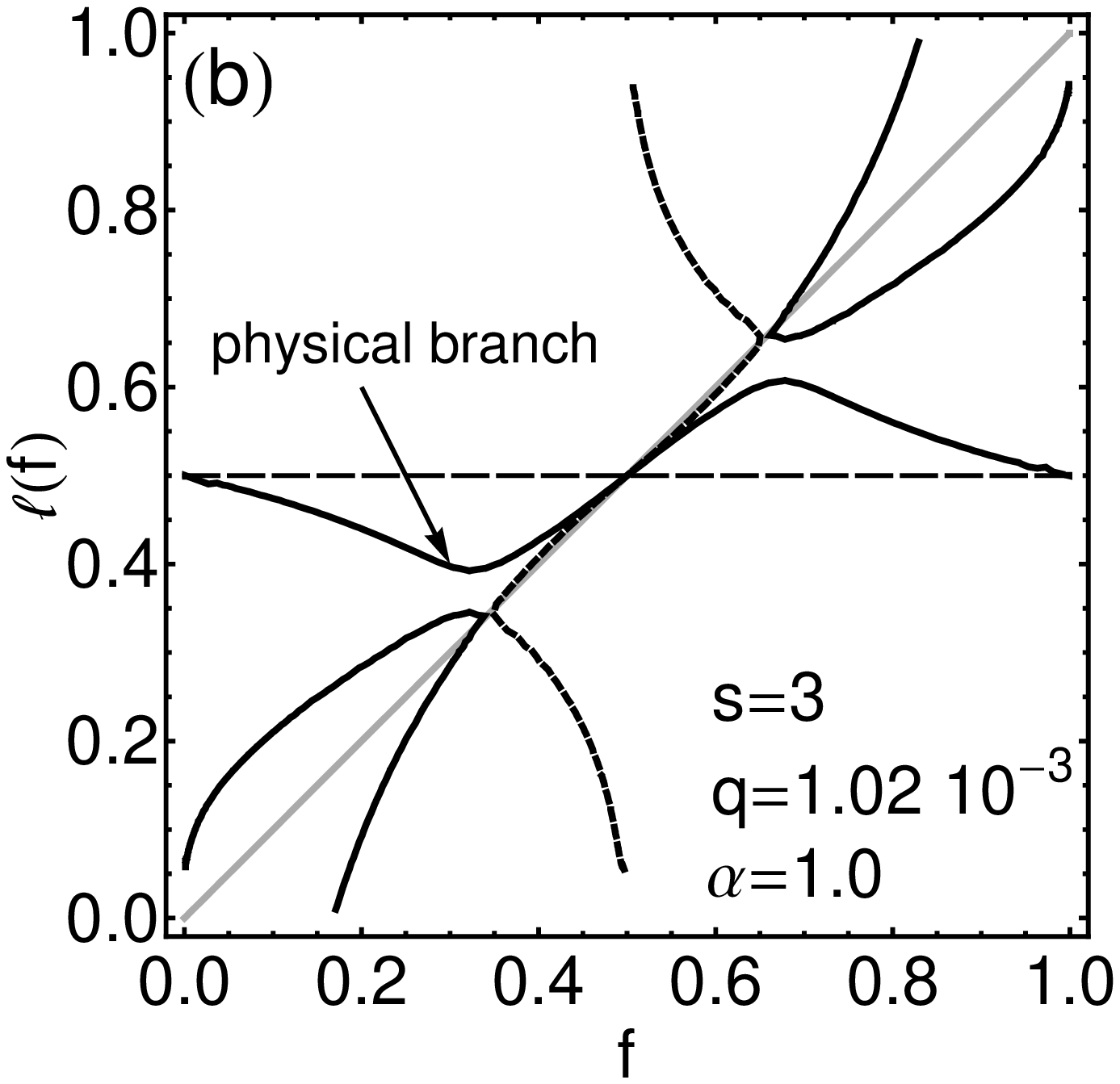}
\caption{\label{fig:fig5} Dimension $d=2$ and contrast $10^8$. Typical conformation of the branches of $\ell$ for $d=2$ with (a) a `bad' case; (b) a `good' one (see text).}
\end{figure}

Depending on parameters $\{f,s,q,\alpha,\beta\}$ the theory admits up to three real solutions. Admissible values of $\{s,q,\alpha,\beta\}$ must be such that a continuous real function $\ell(f)$ exists \emph{for all contrasts} in the interval $f\in(0,1)$, with endpoint values $\ell=1/d$. We call it the ``physical branch''. A criterion that generalizes (\ref{eq:constrs}) to the whole parametric domain seems out of reach. Still, by requiring real solutions for positive permittivities and any $0\leq\ell\leq 1$, an argument similar to the one used in Sec.\ \ref{sec:gmgt} leads to the constraint
\begin{equation}
\label{eq:constrsq}
q s^d>(d-1),
\end{equation}
which extends (\ref{eq:constrs}) to values of $q\not=1$. The actual permitted domain depends on $(\alpha,\beta)$ and is certainly wider, since solutions $\ell$ take on values in a much more restricted interval. Thus, the above constraint is too restrictive in practice.

Numerical experimentations show that for $d=3$, percolation thresholds markedly lower than $1/3$ are obtained for parameter values close to the boundary of the domain, much as in the simpler case of the previous Section, illustrated by expression (\ref{eq:percoab}) of the threshold. As a rule, close to boundary, the smaller $q$, the larger $s$, which is consistent with Eq.\ (\ref{eq:constrsq}) and with the interpretation of Fig.\ \ref{fig:fig2}: the more dispersed the clusters, the larger their effective radius. But for lack of any simple analytical expression of the boundary of the allowed domain, no systematic study of the threshold in $d=3$ was made.

Fig.\ \ref{fig:fig5} displays some typical conformation patterns of solution branches of $\ell$ for $d=2$. Parameters, as indicated  (recall that $\beta$ is irrelevant for $d=2$), were chosen such that $\ell'(f=1/2)\simeq 1$. The expression of $\ell'(f=1/2)$ is easily deduced by series expansions. The solid (resp.\, dashed) lines represent solutions arising from using the `plus' (resp.\ `minus') branch in Eq.\ (\ref{eq:ebsols}). Lines interrupt themselves where solutions become imaginary. Bad parameter values, such as in Fig.\ \ref{fig:fig5}(a), lead to discontinuous real solutions $\ell(f)$. Fig.\ \ref{fig:fig5}(b) illustrates a ``good' case, with the physical branch indicated. Although lying close to $f$ in a region around $f_c=1/2$, the function $\ell(f)$ could not be made close enough to reproduce the double-threshold effect of checkerboards mentioned in Sec.\ \ref{sec:prelrem}. The physical branch was always found to be generated from the `plus' solution of (\ref{eq:ebsols}) in all cases examined where it exists. Whereas in the infinite-contrast limit and for some special parameter sets, joining `plus' and `minus' branches can produce a composite continuous $\ell(f)$ graph with endpoints $1/d$, such special solutions do not survive at lower contrasts, and cannot be considered physical in the present context (although the situation might change upon including three-body interactions).

With $\zeta_2$ as in (\ref{eq:zet2}), and introducing
\begin{equation}
\alpha_q=[q+(1-q)\alpha]/q,\quad \beta_q=[q+(1-q)\beta]/q,
\end{equation}
and the function $g_2(z)\equiv dz-g(z)$, the dilute expansion reads
\begin{equation}
\ell_e^{\rm sc\ell}(f)=\frac{1}{d}+\frac{d-1}{d^2}\left[\alpha_q g_2\left(\frac{\zeta_2}{s^d}\right)-\beta_q h\left(\frac{\zeta_2}{s^d}\right)\right]f.
\end{equation}
When $q=1$, it reduces to (\ref{eq:lsc}) thanks to identity (\ref{eq:gph}). The infinite-contrast slopes at $f=0$, $1$ become:
\begin{subequations}
\begin{eqnarray}
\label{eq:sig0scell1}
\sigma_0^{\rm sc\ell}&=&\frac{d-1}{d^2}\left[\alpha_q g_2\left(s^{-d}\right)-\beta_q h\left(s^{-d}\right)\right],\\
\label{eq:sig1scell1}
\sigma_1^{\rm sc\ell}&=&\frac{d-1}{d^2}\left[\alpha_q g_2\left(\frac{s^{-d}}{d-1}\right)+\beta_q h\left(\frac{s^{-d}}{d-1}\right)\right].
\end{eqnarray}
\end{subequations}
\begin{figure}[!ht]
\centering
\includegraphics[width=5.5cm]{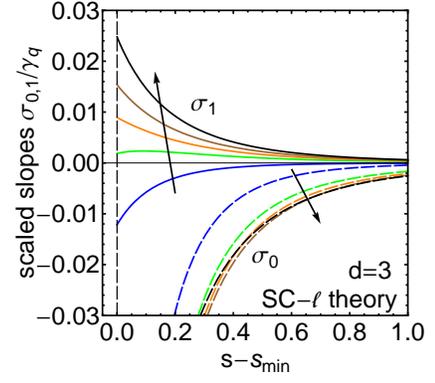}
\caption{\label{fig:fig6} (Color online) Dimension $d=3$. Scaled slopes vs.\ parameter $s$, in s.c.\ theory with variable $\ell$, for $2\phi/\pi=0.0$, $1/3$, $1/2$, $2/3$, and $1$. Dashed: $\sigma_0$; solid: $\sigma_1$. Arrows indicate the direction of variation as $\phi$ increases (color online).}
\end{figure}
With $q\not=1$, the slopes now depend on $\alpha$ and $\beta$, with an overall scaling by $1/q$. The effect of varying  $\alpha_q$ and $\beta_q$ can be studied by factoring out $\gamma_q\equiv(\alpha_q^2+\beta_q^2)^{1/2}$, and by introducing the angle $0\leq\phi\leq\pi/2$ such that $\phi=\arctan(\beta_q/\alpha_q)$, for various values of $\phi$.

In two dimensions the common slope is $\alpha_q$ times that of Fig.\ \ref{fig:fig3}(b). Fig.\ \ref{fig:fig6} displays graphs of the scaled slopes $\sigma_{0,1}^{\rm sc\ell}/\gamma_q$ vs.\ $s$ in three dimensions for some values of $\phi$. The important point is that values of $\sigma_1$ with either sign are now available. When $s^d\gg 1$ and $q\ll 1$ the slopes behave asymptotically as
\begin{subequations}
\begin{eqnarray}
\label{eq:sig0asympt2}
\sigma_0^{\rm sc\ell}&=&\sigma_1^{\rm sc\ell}\sim -\alpha/(6q s^6)\quad\text{if}\quad d=2,\\
\sigma_0^{\rm sc\ell}&\sim&
\left\{
\begin{array}{rl}
\displaystyle{-\beta/(3q s^6)}&\text{if}\quad d=3, \beta\not=0\\
\displaystyle{-2\alpha/(3q s^9)}&\text{if}\quad d=3, \beta=0,
\end{array}
\right.\\
\sigma_1^{\rm sc\ell}&\sim&
\left\{
\begin{array}{rl}
\displaystyle{\beta/(12q s^6)}&\text{if}\quad d=3, \beta\not=0\\
\displaystyle{-\alpha/(12q s^9)}&\text{if}\quad d=3, \beta=0.
\end{array}
\right.
\end{eqnarray}
\end{subequations}

\begin{figure*}[!ht]
\centering
\includegraphics[width=4.3cm]{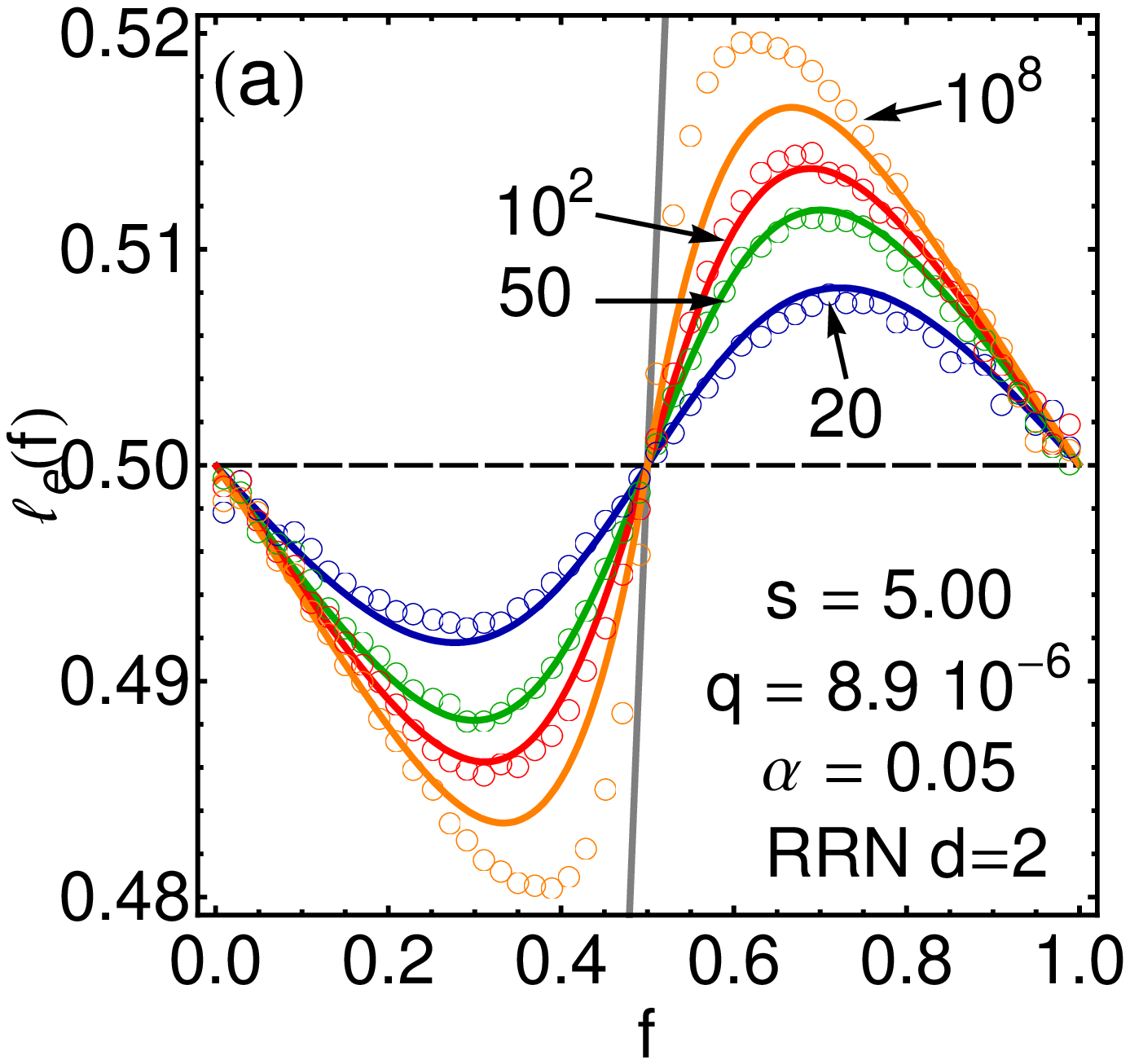}
\includegraphics[width=4.3cm]{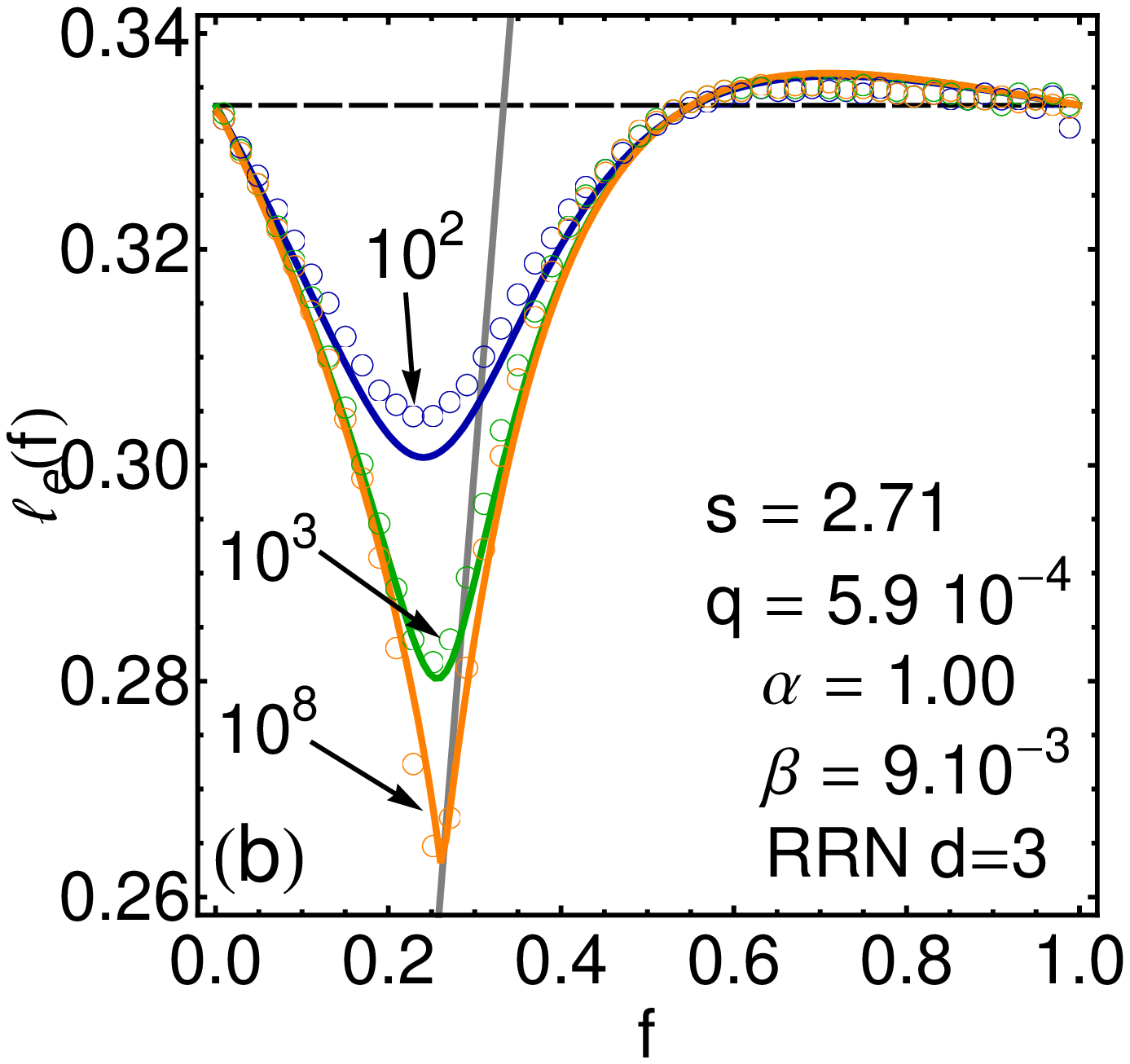}
\includegraphics[width=4.3cm]{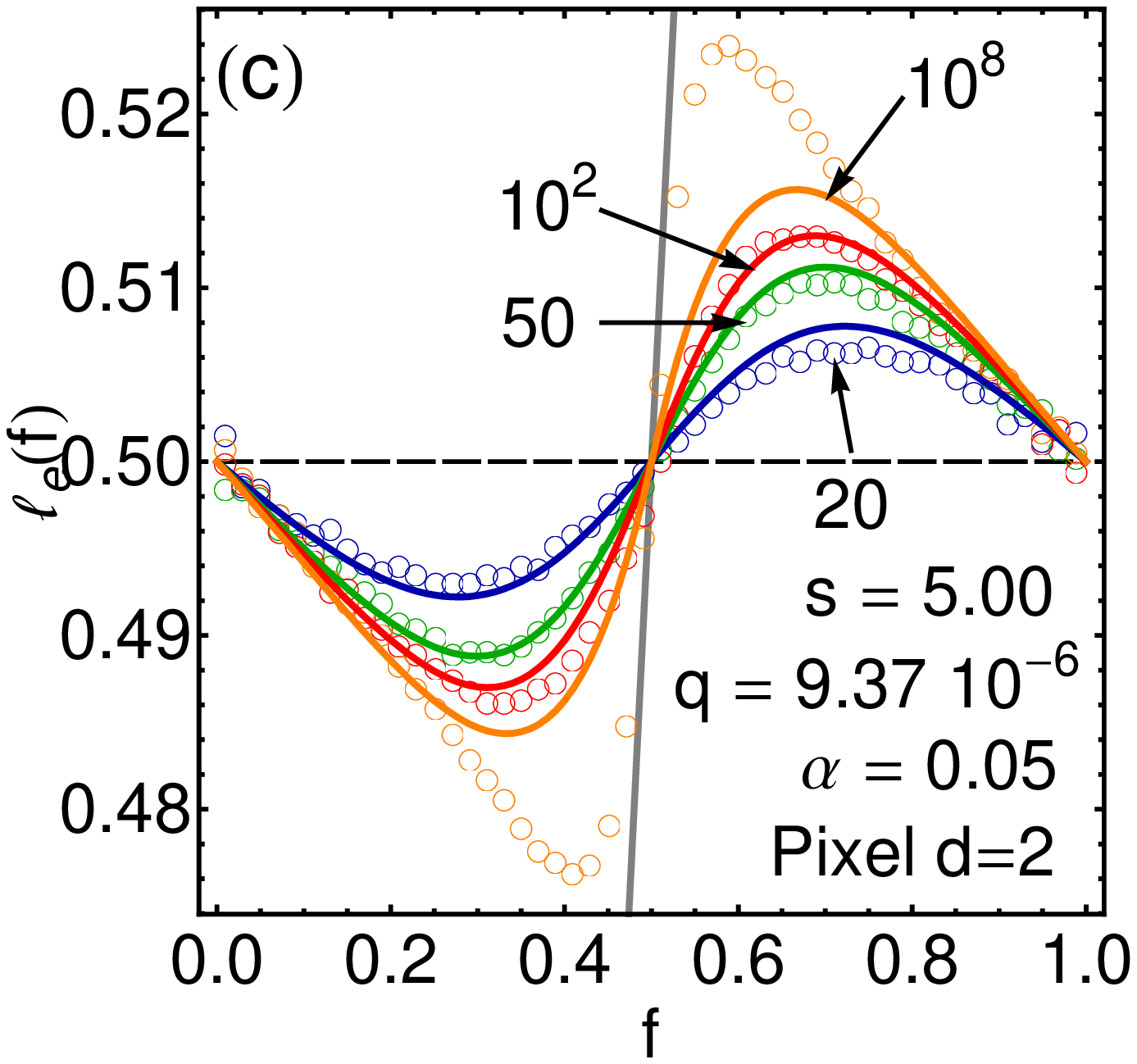}
\includegraphics[width=4.3cm]{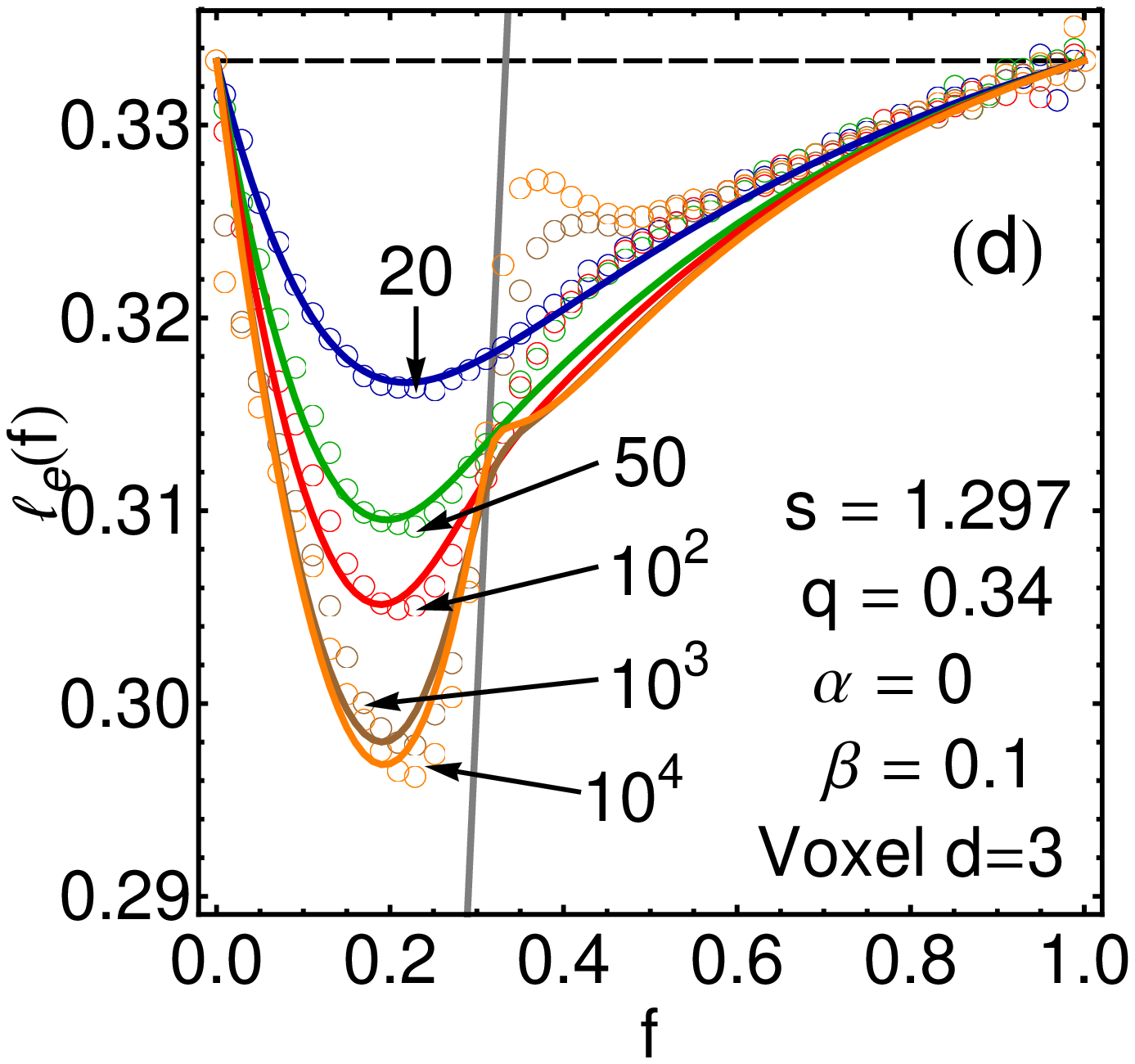}
\caption{\label{fig:fig7} (Color online) Theoretical plots of $\ell_e(f)$ obtained from the s.c.\ theory with variable $\ell$ (solid) fitted on the data of Fig.\ \ref{fig:fig1} (dots). Contrasts and parameters values are as indicated.}
\end{figure*}

Figure \ref{fig:fig7} compares the simulation data of Fig.\ \ref{fig:fig1} with theoretical plots drawn using suitable parameter values in the s.c.\ equations. Parameter values were constrained by imposing the theoretical slopes at $f=0,1$ to match on average those read from the data, and by requiring a high slope at $f=1/2$ in the two-dimensional case, which is achieved by taking $s$ arbitrary, but large. This does not uniquely determine the parameters, so some variations are admissible.

In both two-dimensional cases of Figs.\ \ref{fig:fig7}(a) and (c), good agreement with the data is obtained for low to moderate contrasts, but the high-contrast data values for $f$ close to $f_c$ cannot be matched. Actually, the peak values of $\ell_e(f)$ at high contrast could have been reached by using other parameters, but at wrong volume fractions and at the expense of the low-contrast fits. It can be shown that for $d=2$ the theoretical slope at $f=1/2$ is at most $\sigma(f=1/2)=-4\sigma(f=0,1)$, which is obtained for $s$ large. This explains why the high-contrast data cannot be matched. Indeed, $\sigma(f=1/2)$ should ideally be equal to $1$, whereas $\sigma(f=0,1)\simeq -0.05$ in both sets of two-dimensional data.

The best overall fit is obtained in the $d=3$ RRN case, Fig.\ \ref{fig:fig7}(b). Good match is obtained in the three-dimensional PDA case of Fig.\ \ref{fig:fig7}(d) also, but in the low-permittivity region ($f<f_c$) only. In the high-permittivity region ($f>f_c$), the dilute slope is well reproduced, and a cusp near $f_c^+$ at high contrast is retrieved; however, $\ell_e(f)$ is too low, and cannot be turned into matching the data by varying parameters. This case requires a $q$ value much larger than for the other three and, correspondingly, a relatively small $s$ value.

\section{Concluding discussion}
\label{sec:concl}
We summarize our findings, and discuss possible improvements and directions for further work. First, we proposed to interpret effective-permittivity data in terms of an overall depolarization-coefficient function $\ell_e$, which proved a sensitive probe to highlight differences between models in a neat way, for all concentrations and contrasts.

Our theory rests on the introduction of an inner free depolarization coefficient $\ell$. Together with the usual free background permittivity, $\varepsilon_b$, these quantities are determined by coupled s.c.\ equations: a Bruggeman-like equation on the polarizatibility of clusters, and a CPA-like equation on the overall self-energy. They reduces to Bruggeman's with $\ell=1/d$ in the absence of many-body corrections. The possibility of introducing $\ell$ and $\varepsilon_b$ stems from an overall invariance property of multiple-scattering theory under such parametrizations. Although we assumed $\varepsilon_b$ and $\ell$ to be scalars, it is clear that the most general reparametrization should involve tensors. This possibility was not considered, but would be necessary, e.g., to compute field fluctuations, since this involves considering anisotropic perturbations to the permittivities \cite{BERG78,*PELL00b}. As it stands, the theory only applies to cases where the one-body term in the effective permittivity is isotropic in the dilute limit, given by Eq.\ (\ref{exactdof}).

Two empirical parameters $s$ and $q$ were introduced to describe clusters. While values $\ell\not=1/d$ suggests that the relevant clusters of the theory are not spherical, computations with two-body terms were carried out with a spherical exclusion volume (parametrized by $s$), for simplicity; we note that a parameter $n$, similar to our $s$, was previously introduced by Cichocki and Felderhof in two-body integrals to the purpose of studying scaling relationships in corrections to the CM formula \cite{CICH97}.

Allowing for $\ell\not=1/d$ involves products of Dirac functions. The latter arise from a rough treatment of clusters as point-like polarizable entities, within which dipolar interactions occur only through the local part of the Green function. The resulting mathematical ambiguity was handled by introducing $q$, interpreted as a covering parameter between inner inclusions. The main operational role of parameters $s$ and $q$ is to modify the three-dimensional percolation threshold of the theory. Inasmuch as they represent microstructutral features of the medium, these parameters are akin to the geometrical ones introduced by Miller \cite{MILL69a}, although we cannot claim any precise connection at this point.

We moreover needed to introduce two supplementary parameters $\alpha$ and $\beta$ to handle independently the local and non-local two-body parts in the CPA self-consistency condition on the self-energy. When only part of the latter is canceled out, the theory stands as intermediate between CM-like and BL-like approaches. This additional flexibility was required to produce realistic $\ell_e(f)$ functions.

Although our emphasis was on versatility, the present theory could be rigorously reformulated in the discrete framework adapted to RRNs, for which the two-body interaction term is known exactly (without needing to introduce parameters $s$ and $q$) \cite{NIEU87,CLER96,PELL92}, in order to investigate precisely the role of parameters $\alpha$ and $\beta$. We also remark that while the percolation threshold can be adjusted for $d=3$, where it depends on all four parameters, it remains stuck to its exact bond value $f_c=1/2$ in $d=2$. This does dot allow one to handle site percolation \cite{SAHI03}. Improving the behavior near $f_c$, as well as attempting to reproduce site-percolation effects within the present theory, would presumably require including three-body interactions \cite{CICH89b}. It would be possible, if needed, to introduce additional parameters similar to $\alpha$ and $\beta$ when including their contribution in the generalized CPA condition.

Finally, the multiple-scattering formulation in the continuum allows in principle for extensions to elasticity \cite{KORR73}, or to the frequency domain, including wave propagation dynamics \cite{RECH08,*ROLL11}.

\begin{appendix}
\section{$N$-body expansion in multiple-scattering framework}
\label{appA}
To prove (\ref{nord}), we first demonstrate that
\begin{equation}
\label{recur}
T^{(n)}_{(i_1,i_2,\ldots,i_n)}=T^{(n-1)}_{(i_1,i_2,\ldots,i_{n-1})}
G_\ell u_{\ell i_n} \Bigl(1-G_\ell \sum_{p=1}^{n} u_{\ell i_p}\Bigr)^{-1},\\
\end{equation}
for $n\geq 2$, the recursion being initiated with
$T^{(1)}_{(i)}\equiv t_i$. We start from the expression
$\mathcal{T}=\sum_i u_{\ell i}(1-G_\ell \sum_j u_{\ell j})^{-1}$,
obtained from the equivalence between (\ref{g}) and
(\ref{difmul1}), and to which we apply systematic transformations.
Introduce
\begin{equation}
a_i\equiv(1-G_\ell u_{\ell i})^{-1},\quad
A_i\equiv\Bigl(1-G_\ell\sum_{k\neq i} u_{\ell k} a_i\Bigr)^{-1}.
\end{equation}
Then, with $T_{(i)}^{(1)}\equiv t_i=u_{\ell i}a_i$,
\begin{eqnarray}
{\cal T}&=&\sum_i u_{\ell i} \Bigl(1-G_\ell u_{\ell i}-G_\ell
\sum_{j\neq i} u_{\ell j}\Bigr)^{-1}
=\sum_i t_i\,A_i\nonumber\\
&=&\sum_i t_i\Bigl(1+G_\ell\sum_{j\neq i} u_{\ell j}  a_i A_i\bigr)\nonumber\\
\label{step1} &=&\sum_i T_{(i)}^{(1)}+\sum_{\atop{i}{j\neq
i}}T_{(i)}^{(1)}G_\ell u_{\ell j} a_i A_i,
\end{eqnarray}
Let now $a_{ij}\equiv[1-G_\ell (u_{\ell i}+u_{\ell j})]^{-1}$. One has
\begin{eqnarray}
&&A_i^{-1}=1-G_\ell u_{\ell j}\,a_i-G_\ell\sum_{k\not=i,j} u_{\ell k}\,a_i\nonumber\\
&=&\Bigl[1-G_\ell\sum_{k\not=i,j} u_{\ell k} a_i\left(1-G_\ell u_{\ell j} a_i\right)^{-1}\Bigr]
\left(1-G_\ell u_{\ell j} a_i\right)\nonumber\\
\label{transform1}
&=&\Bigl(1-G_\ell\sum_{k\not=i,j}u_{\ell k}
 a_{ij}\Bigr)a_{ij}^{-1} a_i.
\end{eqnarray}
Whence, with
$A_{ij}\equiv\Bigl(1-G_\ell\sum_{k\not=i,j} u_{\ell k}\,a_{ij}\Bigr)^{-1}$, and
\begin{equation}
T_{(i,j)}^{(2)}\equiv T_{(i)}^{(1)}G_\ell u_{\ell j}\,a_{ij},
\end{equation}
it follows that
\begin{eqnarray}
{\cal T}&=&\sum_i T_{(i)}^{(1)}+\sum_{\atop{i}{j\neq
i}}T_{(i,j)}^{(2)}A_{ij}\nonumber\\
&=&\sum_i T_{(i)}^{(1)}+\sum_{\atop{i}{j\neq
i}}T_{(i,j)}^{(2)}+\hspace{-1em}\sum_{\atop{\atop{i}{j\neq i}}{k\not=i,j}}T_{(i,j)}^{(2)}G_\ell
u_{\ell k}a_{ij}\,A_{ij},\nonumber\\
\label{step2}
&&
\end{eqnarray}
The progression from (\ref{step1}) to (\ref{step2})
represents one transformation step. Going on by applying
to the last term of (\ref{step2}) a transformation similar to
(\ref{transform1}), namely,
\begin{equation}
A_{ij}^{-1}=1-G_\ell u_k\,a_{ij}-G_\ell\sum_{l\not=i,j,k} u_{\ell l}\,a_{ij},
\end{equation}
and so on, proves (\ref{recur}). Proving (\ref{nord}) is now simple. Assume
(\ref{nord1}) to hold at rank $n$, and evaluate ${\cal
T}^{(n+1)}_{(i_1 i_2\ldots i_{n+1})}$. Let
$$
B_n=\Bigl(1-G_\ell \sum_{p=1}^{n} u_{\ell i_p}\Bigr)^{-1}.
$$
Because of (\ref{recur}), it suffices to show that
\begin{eqnarray}
\label{todem} &&u_{\ell i_{n+1}}B_{n+1}\\
&=&t_{i_{n+1}} \left(1-G_\ell S_{i_1 i_2\ldots i_{n}}G_\ell
t_{i_{n+1}}\right)^{-1}\left(1+G_\ell S_{i_1 i_2\ldots
i_{n}}\right)\nonumber
\end{eqnarray}
with $S_{i_1 i_2\ldots i_{n}}$ expressed in terms of $S_{i_1
i_2\ldots i_{n-1}}$ as in (\ref{nord2}), knowing that
\begin{eqnarray}
\label{known} u_{\ell i_n} B_n&=&t_{i_n} \left(1-G_\ell S_{i_1
i_2\ldots i_{n-1}}G_\ell
t_{i_n}\right)^{-1}\nonumber\\
&\times&\left(1+G_\ell S_{i_1 i_2\ldots i_{n-1}}\right).
\end{eqnarray}
Since by definition [cf.\ (\ref{ti})] $u_{\ell i_n}=t_{i_n}(1+G_\ell
t_{i_n})^{-1}$, (\ref{known}) implies
\begin{eqnarray}
B_n^{-1}&=&\left(1+G_\ell S_{i_1
i_2\ldots i_{n-1}}\right)^{-1}\nonumber\\
&&\hspace{-3em}\times\left(1-G_\ell S_{i_1 i_2\ldots i_{n-1}}G_\ell
t_{i_n}\right)(1+G_\ell t_{i_n})^{-1}.
\end{eqnarray}
Therefore
\begin{eqnarray}
B_{n+1}^{-1}&=&B_{n}^{-1}-G_\ell t_{i_{n+1}}(1+G_\ell t_{i_{n+1}})^{-1}\nonumber\\
&=&\left(1+G_\ell S_{i_1 i_2\ldots i_{n}}\right)^{-1}\nonumber\\
&&\hspace{-3em}\times\left[1-\left(1+G_\ell S_{i_1 i_2\ldots i_{n}}\right)G_\ell t_{i_{n+1}}(1+G_\ell t_{i_{n+1}})^{-1}\right]\nonumber\\
&=&\left(1+G_\ell S_{i_1 i_2\ldots
i_{n}}\right)^{-1}\nonumber\\
&&\hspace{-3em}\times\left(1-G_\ell S_{i_1 i_2\ldots i_{n}}G_\ell
t_{i_{n+1}}\right)(1+G_\ell t_{i_{n+1}})^{-1},
\end{eqnarray}
where use has been made of (\ref{nord2}) in the second line. The
result follows.

\section{Square of the Dirac distribution}
\label{app:distmult}
One possible definition of distribution products, due to Colombeau \cite{COLO90,GSPO08}, is as equivalence classes, whose representative chosen for calculations must be inferred from the context. This concept allows one to work properly with objects such as $\delta^2$, which is proportional to $\delta$, but with non-standard (infinite) proportionality constant \cite{GSPO08}.

In the one-dimensional case for instance, take $\delta(x)=\lim_{\sigma\to 0}\delta_\sigma(x)$ where $\delta_\sigma(x)=e^{-(x/\sigma)^2/2}/(\sqrt{2\pi}\sigma)$ is a Gaussian delta-sequence. Then $\int d x\,\delta_\sigma^2(x)=1/(\sigma\sqrt{4\pi})$. This allows one to set $\sigma\delta^2(x)\equiv q\delta(x)$ with $q=1/\sqrt{4\pi}$, as $\sigma\to 0$. A different choice of delta-sequence would lead to some other $q$. This quantity therefore depends on the choice of representation of $\delta(x)$, motivated by the physical nature of the problem considered. The $d$-dimensional generalization of this argument leads to (\ref{eq:delta2}).

We note that, in principle, a more fundamental treatment of the $d$-dimensional case would require acknowledging that the Dirac contribution to the Green function (\ref{green}) stems from a representation (using a superscript to emphasize the dimension) \cite{GSPO08}
\begin{equation}
\delta^{(d)}(\mathbf{r})=\lim_{\eta\to 0}\delta^{(d)}_\eta(\mathbf{r}),\quad \delta^{(d)}_\eta(\mathbf{r})=\frac{\delta(r-\eta)}{S_d\, r^{d-1}},
\end{equation}
where $\delta(r-\eta)$, with $\eta$ \emph{identical} to that in the principal value prescription in Eq. (\ref{green}), materializes the surface of the Lorentz cavity associated to the exclusion volume between interacting polarizable elements \cite{VANB61}.  Doing so would provide the $q$ of Eq.\ (\ref{eq:delta2}) as a function of $\eta$, but also inevitably introduce other arbitrary constants, leading to unnecessary complications.

\end{appendix}
\bibliography{mybib}
\end{document}